\documentclass[fleqn,usenatbib]{mnras}

\usepackage{amssymb}
\usepackage{newtxtext,newtxmath}

\usepackage[T1]{fontenc}

\DeclareRobustCommand{\VAN}[3]{#2}
\let\VANthebibliography\thebibliography
\def\thebibliography{\DeclareRobustCommand{\VAN}[3]{##3}\VANthebibliography}

\usepackage{graphicx}	
\usepackage{amsmath}	
\usepackage{xcolor}	

\newcommand{\dd}{\text{d}}

\newcommand{\change}[1]{{\textcolor{black}{#1}}}

\newcommand{\MpcOh}{ \,  \mathrm{Mpc}  \, h^{-1} }

\newcommand{\nn}{ \nonumber }

\newcommand{\kmsmpc}{\>{\rm km}\,{\rm s}^{-1}\,{\rm Mpc}^{-1}}

\usepackage{scalerel}

\newcommand{\beq}{\begin{equation}}
\newcommand{\eeq}{\end{equation}}
\newcommand{\beqa}{\begin{eqnarray}}
\newcommand{\eeqa}{\end{eqnarray}}

\title[Measuring BAO with SC]{ Measurement of the photometric Baryon Acoustic Oscillations with self-calibrated redshift distribution }

\author[R. Song et al.]{
Ruiyu Song,$^{1,2}$
Kwan Chuen Chan,$^{1,2}$\thanks{E-mail: chankc@mail.sysu.edu.cn (KCC)}
Haojie Xu$^{3,4,5}$\thanks{E-mail: haojie.xu@shao.ac.cn (HX)}, 
and Weilun Zheng$^{1,2}$
\\
$^{1}$School of Physics and Astronomy, Sun Yat-Sen University, 2 Daxue Road, Tangjia, Zhuhai, 519082, China \\
$^{2}$CSST Science Center for the Guangdong-Hongkong-Macau Greater Bay Area, SYSU, Zhuhai, 519082, China \\
$^{3}$Shanghai Astronomical Observatory, Chinese Academy of Sciences, Nandan Road 80, Shanghai 200240, China\\
$^{4}$Department of Astronomy, Shanghai Jiao Tong University, Shanghai 200240, China\\
$^{5}$Key Laboratory for Particle Astrophysics and Cosmology (MOE)/Shanghai Key Laboratory for Particle Physics and Cosmology, China\\
}

\date{Accepted XXX. Received YYY; in original form ZZZ}

\pubyear{20XX}

\begin{document}
\label{firstpage}
\pagerange{\pageref{firstpage}--\pageref{lastpage}}
\maketitle

\begin{abstract}
We use a galaxy sample derived from the DECaLS DR9 to measure the Baryonic Acoustic Oscillations (BAO). The magnitude-limited sample consists of 10.6 million galaxies in an area of 4974 deg$^2$ over the redshift range of [0.6, 1].  A key novelty of this work is that the true redshift distribution of the photo-$z$ sample  is derived from the self calibration method, which determines the true redshift distribution using the clustering information of the photometric data alone. Through the angular correlation function in four tomographic bins,  we constrain the BAO scale dilation parameter $\alpha$  to be $1.025\pm 0.033 $, \change{consistent with the fiducial Planck cosmology}. Alternatively,  the ratio between the comoving angular diameter distance and the sound horizon, $D_{\rm M} / r_{\rm s}$ is constrained to be $18.94 \pm 0.61 $ at the effective redshift of 0.749.  We corroborate our results with the true redshift distribution obtained from a weighted spectroscopic sample, finding very good agreement. We have conducted a series of tests to demonstrate the robustness of the measurement.  Our work demonstrates that the self calibration method can effectively constrain the true redshift distribution in cosmological applications, especially in the context of photometric BAO measurement.   
\end{abstract}




\section{Introduction}
\label{sec:Intro}

Baryonic Acoustic Oscillations (BAO) manifest as the imprint of primordial acoustic features within the distribution of large-scale cosmic structures \citep{PeeblesYu1970,SunyaevZeldovich1970}. The formation physics of the BAO is linear and thoroughly understood, as exemplified by works such as  \cite{BondEfstathiou1984, BondEfstathiou1987, HuSugiyama1996, HuSugiyamaSilk1997, Dodelson_2003}. 
During the early stages of the universe, baryons and photons form a tightly coupled plasma, giving rise to the excitation of the acoustic oscillations within it. Subsequent to the cosmic microwave background decoupling, photons free stream and the acoustic patterns are preserved in the baryon distribution.  With the accurate calculation of the sound horizon scale, BAO assumes a prominent role as a standard ruler in cosmology \citep{WeinbergMortonson_etal2013, Aubourg:2014yra}. 
Since its clear detection in the Sloan Digital Sky Survey by \citet{Eisenstein_etal2005} and the 2dF Galaxy Redshift Survey by \citet{Cole_etal2005}, BAO measurements have been replicated across various spectroscopic datasets at different effective redshifts \citep{Gaztanaga:2008xz, Percival_etal2010, Beutler_etal2011, Blake2012_WiggleZ, Anderson_BOSS2012, Kazin_etal2014, Ross_etal2015, Alam_etal2017, eBOSS:2020yzd, DESI_BAO_2023}.

On a different note, clustering analyses of imaging datasets are yielding competitive outcomes. Despite the somewhat less precise redshifts (photo-$z$'s), imaging surveys demonstrate efficiency in capturing extensive data volumes with deep magnitude. This attribute proves particularly advantageous for BAO measurements, given its large-scale nature, necessitating substantial datasets for robust statistics.  The uncertainties due to photo-$z$'s are typically as large as the BAO scales, and hence the radial BAO in photometric data cannot be reliably detected. Nonetheless, the transverse BAO can still be measured with confidence \citep{SeoEisenstein_2003, BlakeBridle_2005, Amendola_etal2005, Benitez:2008fs, Zhan:2008jh, Chaves-Montero:2016nmw, Ross:2017emc, Chan:2018gtc, Chan_xiptheory2022, Ishikawa_etal2023}.  A number of photometric BAO measurements have been reported in the literature  \citep{Padmanabhan_etal2007,EstradaSefusattiFrieman2009, Hutsi2010, Seo_etal2012, Carnero_etal2012, deSimoni_etal2013, Abbott:2017wcz, DES:2021esc, Chan_xip2022, DESY6_BAO}. Besides the direct BAO measurement, it is suggested that the reconstruction technique can be used to enhance the strength of the photometric BAO signals as well \citep{Chan_etal2023}.   Currently and in near future, a number of large-scale photometric surveys are in progress, it is conceivable that the photometric BAO will play a more prominent role in future.

Without spectroscopic redshifts (spec-$z$'s), the redshift information of the imaging sample can be derived from the template fitting method, training method, or the clustering-based method (see \citet{Salvato_etal2019,NewmanGruen_2022} for a detailed review). In the template fitting method [e.g.~\citet{Arnouts_1999, Bolzonella_etal2000, Benitez_2000, Ilbert_etal2006}], the photometry data, colors or magnitudes, are fit to a model derived from known SED (Spectral Energy Distribution) templates, with the photo-$z$ as one of the fitting parameter.  The prior information can also be incorporated in the fitting. This method is limited by how accurate and representative the templates are, and also by the accuracy of the prior information.  The training method  is often realized by means of the machine learning approach \citep{CollisterLahav_2004, Sadeh:2015lsa, DeVicente:2015kyp, Zhou_etal2021, LiNapolitano_etal2022}. The training is done via a spec-$z$ sample, and so its accuracy depends on the abundance and the completeness or representativity of the spec-$z$ sample.

 Even if we are already armed with photo-$z$'s of the sample, it is necessary to have the true redshift (true-$z$) distribution of the sample for the modeling.  One can use a matched spec-$z$ sample to determine the true-$z$ distribution. This method crucially depends on the availability of the spec-$z$ sample.  On the other hand, the third approach, the   clustering-based methods can provide an independent way to calibrate the true-$z$ distribution of the photo-$z$ sample. Unlike the previous methods,  this one generally can only constrain the distribution of a group of galaxies. While the previous methods relies on the photometry information, the clustering based method makes use of the clustering information, which is largely due to gravity.  This method assumes that the clustering exists only when the samples spatially overlap.   Depending on the usage of the external spec-$z$ sample, it can be further divided into two types. In the first type, often called clustering-$z$, if there is a spec-$z$ sample that physically overlaps with the photometric sample, by using the angular correlation between them, the true-$z$ distribution of the sample can be determined \citep{ Newman_2008, MatthewsNewman_2010, McQuinnWhite_2013, Menard_etal2013, Morrison_etal2017, Gatti_etal2018, Gatti_etal2022, Cawthon_etal2022}.  The second type, called self calibration (SC), relies solely on the clustering information of the photometric sample itself \citep{Schneider_etal2006, Zhang_etal2010, Benjamine_etal2010, Benjamine_etal2013, Zhang_etal2017, Peng_etal2022, XU2023}.  Compared to clustering-$z$, SC is less used so far, perhaps because the redshift range explored in current surveys are still relatively low ($z\lesssim 1$), and hence the spec-$z$ sample is still sufficient for the calibration purpose. Even in current surveys, the high redshift end of the sample cannot be calibrated by clustering-$z$ method due to lack of spec-$z$ galaxies in high redshift e.g.~\citet{Rau_etal2023, DESY6_BAO}. Thus, we anticipate that the SC method will play a more important role in forthcoming surveys.

In this work we aim to measure the photometric BAO using the SC true-$z$ distribution. Previous works on the application of SC are more orientated towards weak lensing measurements \citep{Erben_etal2009, Benjamine_etal2010, Benjamine_etal2013, Peng_etal2022, XU2023}.  Although the photometric BAO has been measured many times, as far as we know, this is the first time that it is performed on the SC calibrated distribution. The rest of the paper is organised as follows. In Sec.~\ref{sec:Data_sample}, we introduce the data sample used, highlighting its systematics treatment. We describe the calibration of the true-$z$ distribution by the SC method in Sec.~\ref{sec:truez_calibration}. We review the BAO analysis pipeline used to extract the BAO scale, in particular the template and Gaussian covariance, in Sec.~\ref{sec:BAO_analysis_pipeline}. Sec.~\ref{sec:results} is devoted to the BAO fit results and the robustness tests. We conclude in Sec.~\ref{sec:conclusions}.  Throughout this paper, the fiducial cosmology is taken to be  the flat $\Lambda$CDM in Planck cosmology \citep{Planck2020} with $\Omega_{\rm m} = 0.315$, $\Omega_{\Lambda} = 0.685$, $n_{\rm s}=0.965$ and $h=H_0/(100 \kmsmpc) = 0.674$.

\section{Data sample } 
\label{sec:Data_sample}

In this section, we introduce the galaxy sample used for our BAO measurement. In particular, we describe the image systematics treatment performed on the sample. We will also highlight the calibration of the true-$z$ distribution by the SC technique.  The data sample is similar to the one presented in \citet{XU2023}, and we only briefly describe it here and refer the readers to this work for more details.

\begin{figure}
    \centering
    \includegraphics[width=\linewidth]{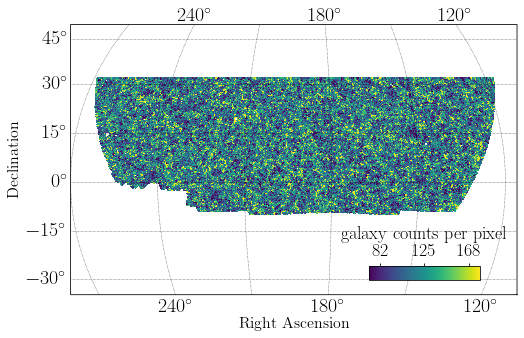}
    \caption{
    Galaxy surface density in the four tomographic bins that we use for BAO fit (Section~\ref{sec:results}).
    The color code is the weighted galaxy counts per {\sc Healpix} \citep{Healpix} pixel ($N_{\rm side}=256$, with the resolution of $\sim 0.23$ deg). 
    The number counts have taken into account of the imaging weights and fractional observed area.}
    \label{fig:footprint}
\end{figure}

The sample is based on the publicly accessible catalog, Photometric Redshifts Estimation for the Legacy Surveys [PRLS, \cite{Zhou_etal2021}], which is further derived from the Data Release 9 of the DESI Legacy Imaging Surveys [LS DR9, \citet{Dey_etal2019}].   LS DR9 combines data from three different surveys: Beijing–Arizona Sky Survey (BASS), Mayall z-band Legacy Survey (MzLS), and Dark Energy Camera Legacy Survey (DECaLS). The imaging dataset consists of data in three  optical bands: g, r, and i.  In addition to these optical bands,  the photo-$z$ estimation process also incorporates two mid-infrared band data [W1 (3.4 $\mu$m) and  W2 (4.6 $\mu$m)] from the Wide-field Infrared Survey Explorer (WISE) satellite. Following \citet{XU2023}, we only consider data in the DECaLS-NGC region, and its footprint is shown in Fig.~\ref{fig:footprint}.  The footprint area is $\sim 4974$ deg$^2$, \change{which includes the veto masks and other selections mentioned below. }

We select the sample following the steps similar to \citet{Yang_etal2021}.  In brief, we first identify extended imaging objects based on the morphological classifications provided by the {\sc TRACTOR} software \citep{Lang_etal2016}. To ensure a robust photo-$z$ estimation, we specifically choose objects with at least one exposure in each optical band. Additionally, we exclude objects with galactic latitude $|b| < 25.0 $ degree to steer clear of high stellar density regions. Lastly, we eliminate objects whose fluxes may be influenced by bright stars, large galaxies, or globular clusters \change{(e.g. the LS DR9 maskbits 1, 5, 6, 7, 8, 9, 11, 12, 13 \footnote{\url{https://www.legacysurvey.org/dr9/bitmasks/}})}. We also apply the same selections to the \change{ publicly available random catalogs \footnote{ \url{https://www.legacysurvey.org/dr9/files/\#random-catalogs-randoms}}. }

\begin{figure}
    \centering
    \includegraphics[width=\linewidth]{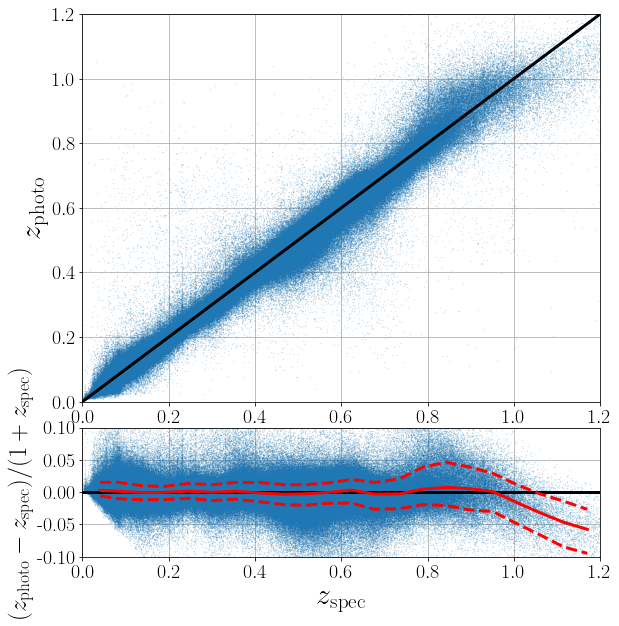}
    \caption{
Photometric redshift performance of the full sample that used to perform the  SC.  
{\it Upper panel}: The $z_{\rm photo}$ versus $z_{\rm spec}$ for
$\sim 1$ million (blue dots) out of 40.5 million galaxies with accurate redshift. The solid black line is the 1:1 line. 
{\it Lower panel}: The fractional difference between $z_{\rm photo}$ and $z_{\rm spec}$ as a
function of $z_{\rm spec}$. The solid (dashed) red line is the median (68 percentile).
The $\sigma_{\rm NMAD}$ and outlier rate are $\sim 0.013$ and $\sim 0.8$\%, respectively.
    }
    \label{fig:specz_photoz}
\end{figure}

We use the photo-$z$'s provided by PRLS, which are derived using the random forest method. Random forest is a machine learning method, and it is trained using secure redshifts from various spectroscopic surveys. After crossing matching to the PRLS sources and homogenizing the sample to avoid bias, about 0.68 million spec-$z$ galaxies were used in training, with 60.1\% from BOSS and SDSS.  \citet{Zhou_etal2021}  demonstrated that the photo-$z$ performance for galaxies with $z_{\rm mag}$ < 21 is reasonably good, reaching a $\sigma_{\rm NMAD} \sim 0.013$ and outlier rate $\sim $1.51\%   \footnote{ $\sigma_{\rm NMAD} \equiv 1.48 \times {\rm median}[ |\Delta z | / ( 1 + z_{\rm spec})]$ and the galaxy is classified as outlier if $|\Delta z | > 0.1 ( 1 + z_{\rm spec}) $. } (see their Figure B1 and B2).  We therefore set the limiting z-band magnitude for our galaxy selection to be $z_{\rm mag} <21 $ for better photo-$z$ performance. \change{  To highlight the photo-$z$ quality of the sample used for SC calibration, in Fig.~\ref{fig:specz_photoz}, we have compared the photo-$z $ with spec-$z$ using a sample of galaxies with spec-$z$ available.  }

Survey properties such as stellar contamination, galactic extinction, sky brightness, seeing, etc can induce artificial variations in the observed galaxy density,  causing spurious signals if not accounted for, especially on large scales \citep{Scranton_etal2002, Ross_etal2011, Ho_etal2012, Ross_etal2017, Rezaie_etal2020, Rodriguez-Monroy_etal2022}.  Therefore, we want to apply some weight on each galaxy to eliminate the impacts of the survey properties.  It is convenient to talk about mitigation in terms of the field (galaxy number or survey properties) values at the pixel level. Since the contamination can only depend on pixel through the value of the survey properties at that pixel,  we can estimate the relationship between the contamination level and the survey property values by fitting (or regression in statistics parlance) to all pixel data available.  In the simplest scenario, we can perform a linear hyperplane fitting in survey property space.  We correct the  systematics contamination employing the random forest mitigation technique developed in \citet{Chaussidon_etal2022}, which has been conveniently encapsulated in the public code {\sc regressis}.  We can effectively think of the random forest method as a sophisticated way to carry out fitting in the multi-dimensional survey property space.

Here, we mention the key information on the mitigation procedure at work. We refer the interested readers for more details to \citet{Chaussidon_etal2022, XU2023}.  The survey property maps we adopt for systematics mitigation are the following: stellar density \citep{Gaia_2018}, galactic extinction \citep{Schlegel_etal1998, Schlafly_etal2011}, sky brightness (g/r/z bands), air mass (g/r/z), exposure time (g/r/z), PSF size (g/r/z), and PSF depth (g/r/z/W1/W2). Altogether, we consider 19 imaging maps.   The mitigation process is performed at the {\sc healpix} \citep{Healpix} pixel level with $N_{\rm side}=256$ ($\sim 0.23$ deg). Note that we choose to only use the high quality pixels \footnote{We only consider pixels with $f_{{\rm pix}, i} > 0.8$, where $f_{{\rm pix}, i}$ is the fractional observed area of pixel $i$ calculated as the ratio of random points after the sample selection and before the selection. } to do the fit.  After quality cuts, there are $\sim$94838 valid pixels in our footprint. 
To avoid the overfitting problem, following the advice in \citet{Chaussidon_etal2022}, we divided the entire footprint into 7 folds (each is about 830 ${\rm deg}^2$), and used 6 folds for training and the remaining one for validation. The hyperparameters for the random forest are the same as in \citet{Chaussidon_etal2022}. The outcome of the random forest mitigation method is a weight map, one for each pixel. This means that galaxies in the same pixel share the same weight.   The procedure is applied to each tomographic bin separately, and so there is a weight map for each bin. A crucial diagnostic of the effectiveness of the procedure is that, after mitigation, the galaxy number density should not show any dependence on the survey properties. Fig.~3 in \citet{XU2023} indeed demonstrates that such a dependence is substantially reduced after mitigation. Moreover, mitigation should also significantly decrease the cross correlation between the galaxy densities and the survey property maps. This also holds, as demonstrated in Fig.~4 in  \citet{XU2023}.

 Table~\ref{tab:data_property} summarizes the properties of our data sample for BAO fit. For the BAO analysis, we have divided the data in the photo-$z$ range [0.6, 1] into four tomographic bins, each of bin width $ \Delta z = 0.1 $.  There are altogether 10,576,969 galaxies after systematics correction. The spatial number density of the whole sample \change{(in the redshift range of $0.6 < z_{\rm ph} <1)$} is  $2.47 \times 10^{-3} \, (\MpcOh)^{-3} $.  We measure the galaxy bias by fitting a theory template (described in details in Sec.~\ref{sec:BAO_template}) to the angular correlation function measurement in the angular range of [0.5, 3] degree.  The effective redshift of the sample is $ z_{\rm eff} = 0.749 $, which is computed using the photo-$z$  $ z_{\rm ph}$ of galaxies in the sample as
\beq
z_{\rm eff} = \frac{ \sum_i w_{ {\rm sys}, i}  z_{\rm ph} }{  \sum_j w_{ {\rm sys}, j}  },
\eeq
where $w_{\rm sys }$ denotes the systematic weight.

We are ready to measure the angular correlation function $w$, which will be used for both the SC and the BAO measurement. For SC, we need both auto and cross bin correlation functions, while for BAO, we only use the auto ones. The correlation function is measured using the Landy-Szalay estimator \citep{LandySzalay_1993}, schematically  
\beq
w_{ij} (\theta ) = \frac{D_iD_{j} - D_i R_j -D_j R_i + R_i R_j }{R_i R_j} , 
\eeq
where $DD$, $DR$, and $RR$ denote the normalized data-data, data-random, and random-random pair counts, respectively. In the pair counting, we have incorporated the imaging mitigation weights. See  \citet{XU2023} for more details. The correlation measurements are conducted with the code {\sc corrfunc} \citep{SinhaGarrison_2020}.

\begin{table}
\centering
\caption{ Properties of the galaxy sample used for BAO fit. The number of galaxies after systematic mitigation, spatial number density, and galaxy bias of the four tomographic bins are shown. }
\begin{tabular}{lccccc}
  \hline
  \hline
   \bf{Redshift}& \textbf{Weighted} $ \bf{N}_\text{gal} $ & $n_\text{gal}/(\text{Mpc}  h^{-1})^{-3}$ & \bf{Galaxy bias} \\ 
  \hline
    $(0.6,0.7)$&4,477,109&$5.276\times10^{-3}$ &1.303  \\ 
    $(0.7,0.8)$&2,729,689&$2.713\times10^{-3}$&1.728   \\ 
    $(0.8,0.9)$&2,103,716&$1.826\times10^{-3}$&1.682   \\ 
    $(0.9,1.0)$&1,266,455&$9.865\times10^{-4}$&1.706    \\ 
    \hline
    \textbf{Total}& 10,576,969& $2.465\times10^{-3}$&- \\ 
    \hline
  \label{tab:data_property}
\end{tabular}
\end{table}

\section{ True-z distribution by self calibration} 

\label{sec:truez_calibration}

Here we discuss the calibration of the true-$z$ distribution, which is crucial to the modeling of the template and covariance to be presented in next section.  We concentrate on SC method, which is the focus of this work.

Because of photo-$z$ uncertainties, the galaxies in a photo-$z$ bin actually come from a number of spec-$z$ bin, i.e.
\beq
 M_{i'} =  \sum_k  Q_{k i'} M_k,
\eeq 
where $M$ is the angular galaxy density,  $ Q_{k i'}$ is the probability that a galaxy in spec-$z$ bin $k$  leaks into the photo-$z$ bin $i'$.  Here we distinguish the photo-$z$ bin from the spec-$z$ one by adding a prime on the photo-$z$ index.

The angular number density correlation between photo-$z$ bins is given by
\begin{align}
\langle M_{i'} M_{j'}  \rangle = \sum_{k,l}  \langle Q_{k i'} M_k Q_{l j'} M_l \rangle . 
\end{align}
By writing $M_{i'} =\bar{M}_{i'}( 1 + \delta_{i'} ) $ and  $M_{i} = \bar{M}_i( 1 + \delta_i ) $ in terms of their mean $\bar{M}$  and fluctuation $\delta$, we get the angular overdensity correlation function 
\begin{align}
\label{eq:C_photoz}
C_{i'j'} = \sum_k P_{ki'} P_{kj'} C_{kk},     
\end{align}
where $ C_{kk} = \langle \delta_k \delta_k \rangle  $ is the spec-$z$ angular overdensity correlation function,  $ P_{ki'} $ is called the scattering rate in \citet{Zhang_etal2010, Zhang_etal2017} and it is related to $ Q_{ki'} $ by 
\beq
P_{ki'} = \frac{  Q_{ki'} \bar{M}_k}{ \bar{M}_{i'} }. 
\eeq
We immediately see that it satisfies the normalization condition 
\beq
\label{eq:P_normalization}
\sum_k P_{ki'} = 1 .
\eeq
In arriving at Eq.~\eqref{eq:C_photoz},  we have assumed that the spec-$z$ bins correlation is non-vanishing only if they fall in the same bin, and this can be shown to be true in Limber approximation [e.g.~\citet{Simon2007}].   Eq.~\eqref{eq:C_photoz} enables us to extract the scattering rate $P_{ki'} $ using the clustering information of the photo-$z$ data alone \citep{Schneider_etal2006, Zhang_etal2010}.

$ P_{ij'}  $ has $ N_{\rm s} N_{\rm p} $ elements, where $ N_{\rm s} $ ($ N_{\rm p} $) denotes the number of spec-$z$ (photo-$z$) bins. After subtracting $ N_{\rm p } $ normalization conditions [Eq.~\eqref{eq:P_normalization}], we have $ ( N_{\rm s} -1 ) N_{\rm p} $ unknowns. If there are $ N_{\theta} $ angular bins in each redshift bin, we have $ N_\theta N_C $ data measurements, where $N_C$ is the number of correlation functions, e.g. $N_C = N_{\rm p} ( N_{\rm p} + 1 )/2  $ if all the auto and cross correlation functions are considered.
To uniquely solve for Eq.~\eqref{eq:C_photoz}, we have $ N_{\rm s} = N_{\rm p}$ by design.
Assuming independence of the measurements, as long as the number of constraints is larger than the number of free parameters, we can solve the overconstrained system, perhaps in the least square sense. 
We note that the assumption of no correlation between neighbouring spec-z bins prevents us from using a too fine spec-$z$ bin width, and this limits the resolution of the $n(z) $ distribution. This is particularly relevant for BAO measurement as we consider relatively thin tomographic bin width. For more discussions on the independence of the measurements, see \citet{XU2023}.

Because Eq.~\eqref{eq:C_photoz} is a system of coupled quadratic equations (cubic if $C_{kk}$  is included) , it is non-trivial to solve. A series of works have been devoted to solving it \citep{Erben_etal2009, Benjamine_etal2010, Benjamine_etal2013, Zhang_etal2017}.   In contrast to previous works, \citet{Zhang_etal2017} solves Eq.~\eqref{eq:C_photoz} in full generality, without resorting to two bin case only \citep{Erben_etal2009} or linear coupling (in $P$) approximation \citep{Benjamine_etal2010}. Here we outline the algorithm in \citet{Zhang_etal2017}, which solves for $P_{ij'} $ (and $C_{ii}$) in two steps. In the first step, starting from an initial guess of $P $, an iterative fixed-point method is applied to look for a preliminary solution. In the absence of noise, this  preliminary solution will be sufficient. But in the presence of noise in realistic situation, the fixed-point solution is easily trapped in one of the shallow local minima. To get closer to the global one, a more power method, the non-negative matrix factorization \citep{LeeSeung1999}, is subsequently used to locate the final solution. Here non-negative matrix means that the matrix has no negative elements. For $P$, it is clearly the case, and for correlation function, if we limit to a small angular range, it also holds. While in \citet{LeeSeung1999}, the non-negative matrix is factorized into two distinct non-negative matrices, for Eq.~\eqref{eq:C_photoz},  \citet{Zhang_etal2017} enforces the equality of the two factors. Using the update rules recommended by \citet{LeeSeung1999}, the solution is solved iteratively using the preliminary solution as the input.  Further improvements in implementation are made in \citet{ Peng_etal2022, XU2023}.  Here we use the public version by \citet{XU2023} \footnote{ \url{https://github.com/alanxuhaojie/self_calibration} } .

\begin{figure*}
    \centering
    \includegraphics[width=7.in]{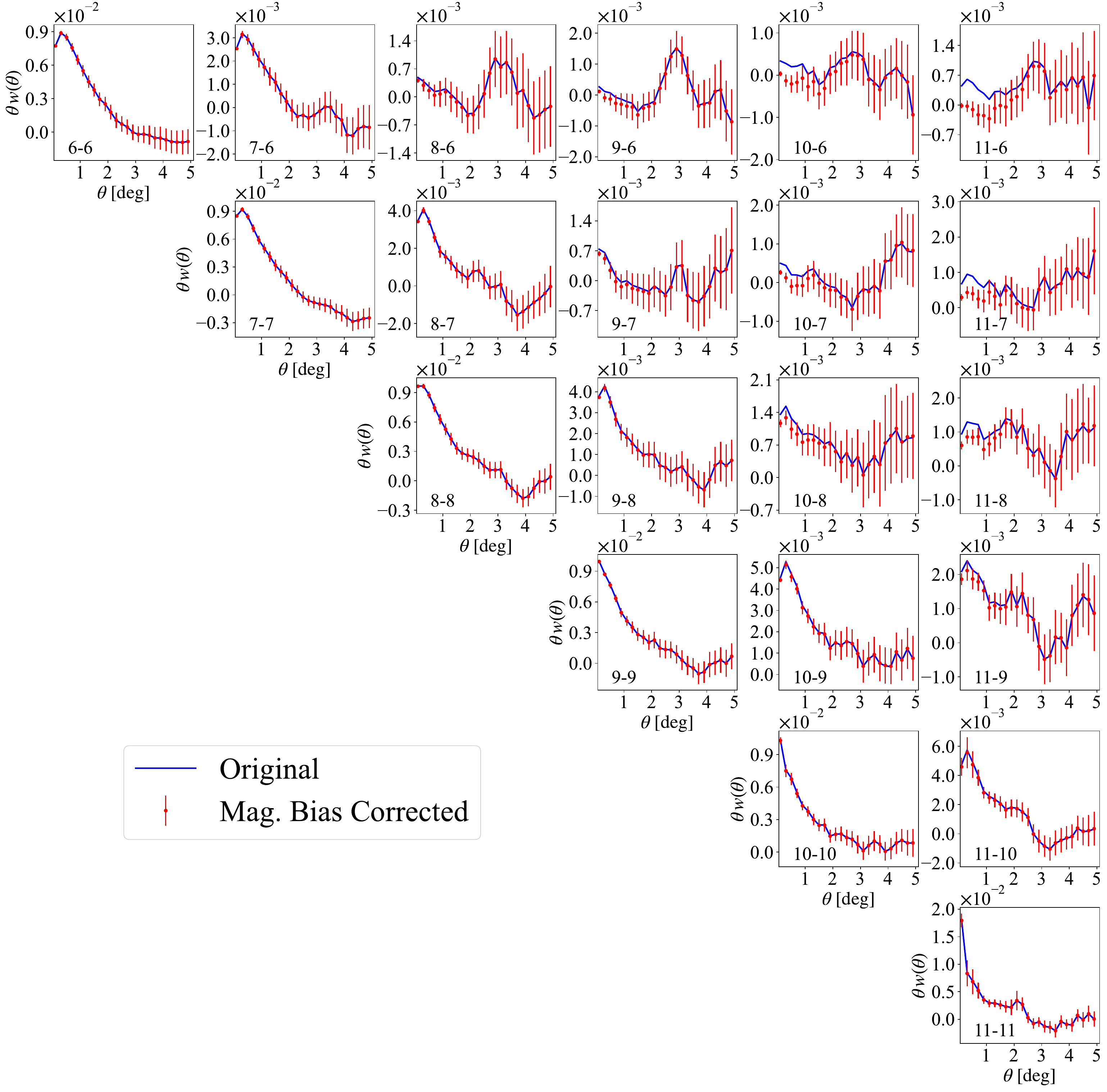}
    \caption{A subsample of the angular correlation function used in SC calibration. The red dots with error bars denote the correlation function with magnification bias corrected, while the blue lines are the original ones without correction. The magnification bias is important for the cross correlation between widely separated bins on small angular scales.  The error bars are given by jackknife re-sampling. To avoid clutter, we only show the results for the bins close to the BAO sample (the redshift range of the BAO sample corresponds to bins 7, 8, 9, and 10). Here, the indices 1, 2, 3, etc denote the photo-$z$ bins (0, 0.1), (0.1, 0.2), (0.2, 0.3), and so on.
    }
    \label{fig:all_correlation}
\end{figure*}

 Even though we will only use the sample in the redshift range [0.6, 1] for BAO measurement, it is important to consider a wider redshift range in SC calibration in order to correctly estimate the true-$z$ distribution.  The SC algorithm is applied to 12 tomographic bins in the redshift range of [0, 1.2], with a $\Delta z$ of 0.1 for each tomographic bin. The angular correlation function is measured in 25 bins of equal angular separation in a linear spacing between 0 and 5 degrees.

 Before the correlation is fed into the SC algorithm, cosmic magnification \citep{MoessnerJain1998, BartelmannSchneider2001} must be removed from the cross-correlation signal between two widely separated tomographic bins. \change{Lensing can increase the area of a patch of the sky, and hence dilute the number density per unit solid angle. On the other hand, it also makes galaxies brighter as lensing conserves surface brightness and consequently, galaxies below the limiting flux can also be observed. The net effect of magnification bias depends on the slope of the luminosity function at the limiting flux.  } To calculate the cosmic magnification, it is necessary to have the true-$z$ distribution of two separate tomographic bins, the linear bias of the near bin, and the slope of the luminosity function of the distant bin at the limiting magnitude. However, none of these are trivially obtained. To proceed, we follow the approximation proposed in \citet{XU2023}, which approximates the true-$z$ distribution by stacking photo-$z$'s convolved with Gaussian photo-$z$ errors from PRLS and estimates the linear galaxy bias from the auto-correlation using the {\sc ccl} \citep{cclpaper}  package assuming the Planck cosmology \citep{Planck2020}. Note that the linear bias is estimated in the range of [0.5, 3] degrees, instead of the full scale used in \citet{XU2023}.  For more information, please refer to \citet{XU2023}. \change{In Sec.~\ref{sec:RobustnessTest}, we shall test the sensitivity of the BAO measurement to these input.} In Fig.~\ref{fig:all_correlation}, we show some of the correlation function input into the SC code. We have applied the imaging systematics mitigation to the measurements. The results with magnification bias subtraction are contrasted with the original ones. We see that the magnification bias is  important for cross correlation between widely separated bins because lensing is efficient when there is sufficient distance between the lens and the source.

\begin{figure*}
    \centering
    \includegraphics[width=5.5in]{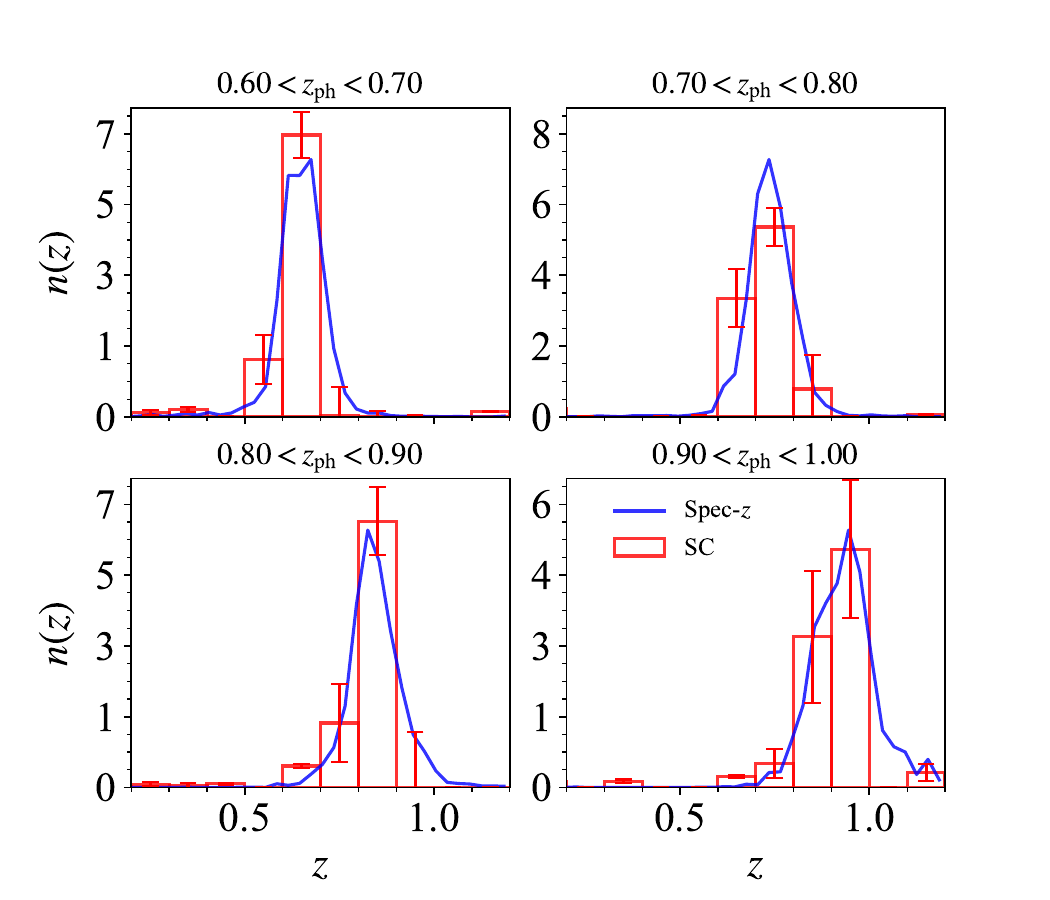}
    \caption{Shown are the true-$z$ distribution $n(z)$ calibrated by the SC (red bars) and weighted spec-$z$ (blue lines) method for the four tomographic bins used for BAO fit. The error bars of the SC distribution are estimated by the NMAD of 100 SC solutions.  }
    \label{fig:SC_z_distribution}
\end{figure*}

As an ancillary to verify the SC results, we consider the weighted spec-$z$ distribution. Because the limiting magnitude in a photo-$z$ survey is often much deeper than that in spec-$z$ sample, the spec-$z$ sample cannot be directly used to calibrate the true-$z$ distribution. This problem can be alleviated by appropriately weighting the spec-$z$ sample to mimick the photo-$z$ sample \citep{Lima_etal2008, Bonnett_etal2016}.

In Fig.~\ref{fig:SC_z_distribution}, we compare the true-$z$ distribution $n(z) $ obtained from SC and weighted spec-$z$.  Note that the redshift resolution of weighted spec-$z$ $n(z)$ \change{($\Delta z = 0.03$)} can be made much finer than the SC case \change{($\Delta z=0.1$)}. As in  \citet{XU2023}, we estimate the best fit and error bars of the SC distribution by generating 100 sets of SC solutions with different initial guesses. We adopt the best fit distribution with the minimal $\chi^2 $, and the error bars are estimated using the NMAD (normalized median absolute deviation) of resultant distributions. NMAD is robust to the outliers of the distribution.  If we adopt the standard deviations as the spread measure instead, the error estimates are generally larger, especially in some of the regions with $n(z) \sim 0 $.  This shows that the distribution of the SC solutions is quite non-Gaussian with strong tails.

In Table \ref{tab:statistics_nz}, we show the summary statistics for the mean and spread of the true-$z$ distribution estimated from SC and weighted spec-$z$. The averages of $n(z)$ estimated by mean $\bar{z}$ and the median of the distribution show similar results. Both suggest that the weighted spec-$z$ yields a mean estimate higher than the SC by $\sim 5 \%$.  For the spread measures, we use the standard deviation $\sigma_{\rm std} $, the NMAD  $\sigma_{\rm NMAD} $, the half width of the central 68 percentile region $\sigma_{68}$. The $\sigma_{\rm std} $ from SC is substantially larger than the weighted spec-$z$ results, because the SC predicts a thicker tail than the spec-$z$. The NMAD and $\sigma_{68}$ show more broadly similar results as they are more robust than $\sigma_{\rm std} $. Except for the second bin, in which the SC gives a much wider spread than spec-$z$, the results are more or less similar for others. With all these differences, however, we shall see that these final constraints on the BAO scale are similar for these two $n(z)$ distributions.

\begin{table}
\caption{ The summary statistics for the $n(z)$ distribution estimated by SC and weighted spec-$z$ for each tomographic bin. Estimates of the average of the $n(z) $ includes the mean and median, and  we show the standard deviation, normalized median absolute deviation and the half width of the central 68 percentile region as estimates of the spread of the distribution.   }
    \centering
    \begin{tabular}{lcccccc}
    \hline
    \hline
        \textbf{photo-}$z$ \textbf{bins} & $\bar{z}$ & $z_\text{median}$ & $\sigma_\text{STD}$ & $\sigma_\text{NMAD}$ &$\sigma_\text{68}$\\

    \hline
    \textbf{SC}\\
    \hline
    (0.6,0.7) &  0.623 & 0.638 & 0.114 & 0.048 & 0.081\\
    (0.7,0.8) &  0.708 & 0.724 & 0.125 & 0.076 & 0.110\\
    (0.8,0.9) &  0.800 & 0.831 & 0.122 & 0.053 & 0.104\\
    (0.9,1.0) &  0.877 & 0.910 & 0.164 & 0.083 & 0.120 \\
    \hline
    \textbf{Spec-$z$}\\
    \hline
    (0.6,0.7) &  0.648  &  0.650 & 0.074 & 0.054 & 0.078\\
    (0.7,0.8) &  0.737  &  0.737 & 0.073 & 0.056 & 0.084\\
    (0.8,0.9) &  0.844  &  0.840 & 0.075 & 0.061 & 0.092\\
    (0.9,1.0) &  0.932  &  0.934 & 0.091 & 0.076 & 0.114\\
    \hline
    \end{tabular}
    
    \label{tab:statistics_nz}
\end{table}

\section{BAO analysis pipeline}
\label{sec:BAO_analysis_pipeline}
With the true-$z$  distribution $n(z)$, we can apply it to compute the theoretical  angular correlation function templates and their covariance, which are the key ingredients to extract the BAO scale.  Our BAO analysis pipeline is essentially the same as \change{the angular correlation function analysis adopted in} DES BAO measurement \citep{Chan:2018gtc, Abbott:2017wcz,DES:2021esc}. \change{ Alternatively, we can apply the angular power spectrum \citep{Camacho_DESY1Cl,DES:2021esc} or the projective correlation function \citep{Ross:2017emc,Chan_xiptheory2022,Chan_xip2022} to extract the BAO signals.  }

\subsection{BAO template}
\label{sec:BAO_template}

The linear matter power spectrum is decomposed into the smooth no-wiggle part, $P_{\text{nw}}$, and the wiggle part, $ P_{\text{lin}}-P_{\text{nw}} $. By modeling the BAO damping effect with a Gaussian damping factor, we have 
\begin{equation}
    P_\text{m}(k, \mu)=[P_{\text{lin}}(k)-P_{\text{nw}}(k)]\text{e}^{-k^2\Sigma^2(\mu)}+P_{\text{nw}}(k), 
    \label{eq:Pk_BAOdamp}
\end{equation}
where the damping scale $\Sigma$ is anisotropic, depending on 
$\mu=\hat{\mathbfit{k}}\cdot\hat{\mathbfit{e}}$, the projection of unit wave vector $\hat{\mathbfit{k}}$ onto the line-of-sight direction $\hat{\mathbfit{e}}$. 
The Gaussian damping factor is computed theoretically using the IR resummation method following \citet{Ivanov:2018gjr}.

The linear redshift-space power spectrum between redshift $z_1$ and $z_2$ reads \citep{Kaiser87}
\begin{equation}
    P(\mathbfit{k},z_1,z_2)= (b_1+f_1\mu^2)(b_2+f_2\mu^2)D(z_1)D(z_2)P_\text{m}(k),
    \label{eq:P_linear}
\end{equation}
where $b$ is the linear bias of the tracer, and $ f \equiv \frac{ \dd\ln{D} }{ \dd\ln{a} }$ is the linear growth rate with $D$  and $a$ being the linear growth factor and the scale factor, respectively. Here $z_1$ and $z_2$ refer to the true redshifts. We get the 3D redshift-space correlation function after a Fourier transform, and the angular correlation function in photo-$z$ space follows from a projection of the 3D one to the angular space:  
\begin{align}
        w(\theta,z_1^{\rm p},z_2^{\rm p}) =  & \int\dd z_1    \int\dd z_2\,n(z_1|z_1^\text{p} )n(z_2|z_2^\text{p} )  \nn \\
         & \times \int \frac{ \dd \mathbfit{k} }{ (2\pi)^3} \,\text{e}^{i\mathbfit{k}\cdot\mathbfit{s} (z_1,z_2,\theta )   }P(\mathbfit{k},z_1,z_2),
    \label{eq:wtheta}
\end{align}
where $z_1^{\rm p}$ ($z_2^{\rm p}$) is the photo-$z$ estimate of the tomographic bin and its corresponding   true-$z$ distribution is $ n(\cdot | z_1^{\rm p}) $ [$n(\cdot | z_2^{\rm p})]$, which can be  obtained from the SC or weighted spec-$z$ method.  In this work, we consider the auto correlation only for BAO fit, and so  $z_1^\text{p} $ and $ z_2^\text{p} $ are equal. \change{In the template computation, we interpolate the $n(z) $ distribution by cubic spline. Note that the cubic spline results in a smooth redshift distribution, but it may also gives rise to small unphysical oscillations at the tail of the distribution where the estimated distribution is close to zero. We do not correct for this as its impact is minimal. }

 To obtain the galaxy bias (shown in Table \ref{tab:data_property}), we first calculate a template assuming isotropic damping in Eq.~\eqref{eq:Pk_BAOdamp} and then fit it to the angular correlation measurement.  The resultant bias parameters are fed into Eq.~\eqref{eq:P_linear} and \eqref{eq:wtheta} to construct the final template. 

\begin{figure}
    \centering
    \includegraphics[width=3in]{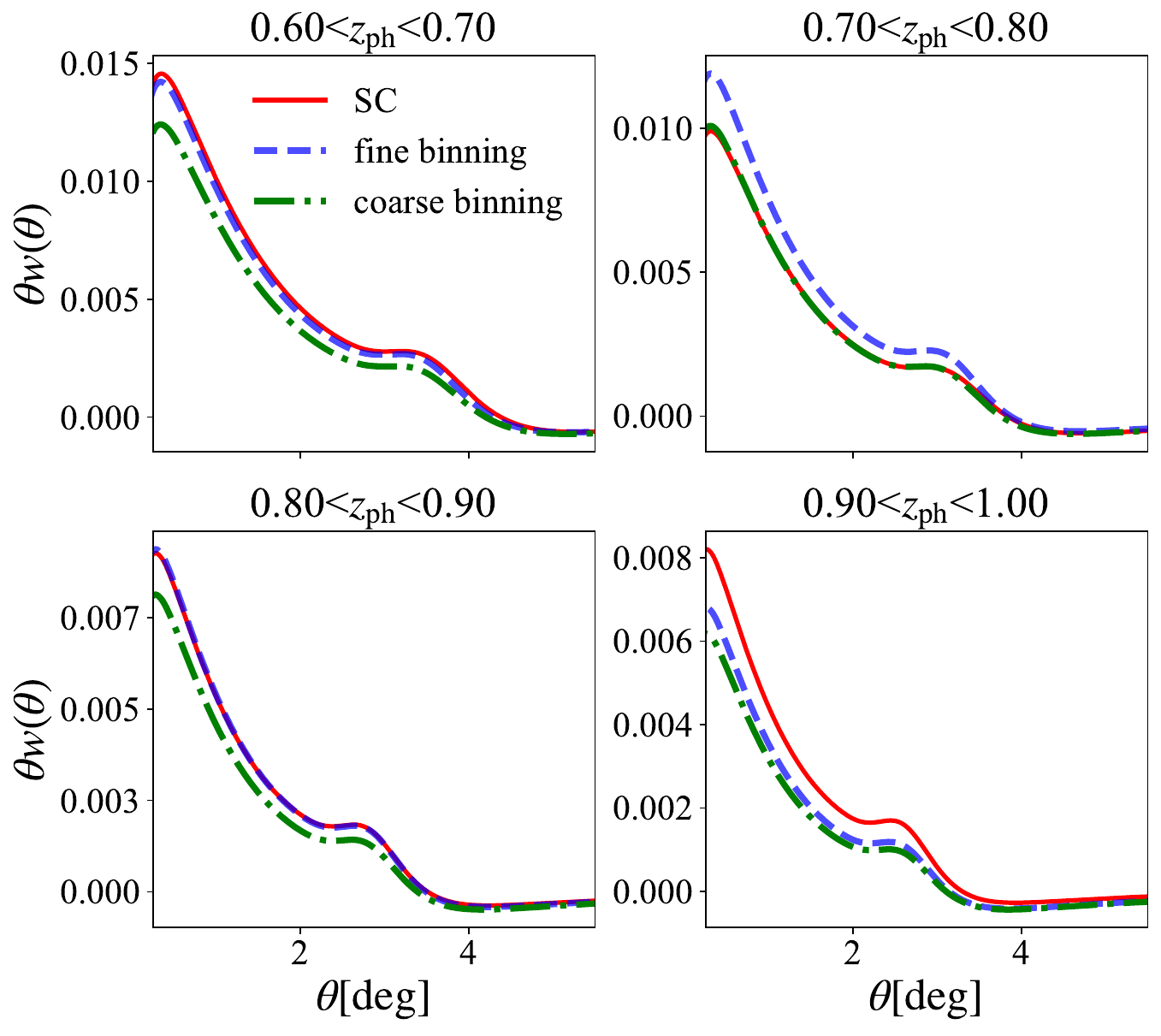}
    \caption{\change{ The templates obtained with different true redshift distributions [SC (solid, red), spec-$z$  $n(z)$ in fiducial fine binning (dashed, blue), and spec-$z$ in coarse binning (dotted-dashed,  green)].  For all the templates shown, the bias parameters are set to be 1.8. While the amplitudes vary slightly between them, the position of the BAO feature remains largely unchanged.  }
  }
    \label{fig:template_diff}
\end{figure}

\change{Fig.~\ref{fig:template_diff} shows the templates obtained from different true-$z$ distribution $n(z)$'s .  We have compared the results from the SC, spec-$z$ in fiducial fine binning, and spec-$z$ in coarse binning. The position of the BAO peak appears to be similar in these cases.  There are small differences in the amplitude of the resultant correlation functions, and they are  mainly controlled by the value of the central "lump" of the $n(z)$ distribution.  We note that this amplitude difference does not affect the BAO measurement and it will be absorbed by the nuisance amplitude parameter in the BAO fit.  }

\subsection{Gaussian covariance matrix}
\label{sec:Gaussian_cov}

As in DES Y3 analyses \citep{DES:2021esc}, our fiducial covariance choice is the analytic Gaussian covariance.  The general Gaussian covariance for the angular correlation function can be written in terms of the  angular power spectrum  $C_\ell $ as \citep{CrocceCabreGazta_2011}
\begin{align}
  \label{eq:cov_mat_crossz}
& \mathrm{Cov} [ \hat{w}_{ij}(\theta),  \hat{w}_{mn}(\theta')] =  \sum_{\ell} \frac{(2 \ell +1)}{(4 \pi)^2 f_{\rm sky}   }   \bar{\mathcal{ L}}_\ell(\cos \theta) \bar{\mathcal{L}}_{\ell}(\cos\theta') \nonumber \\
 & \times \Big[ \Big(  C^{im}_{\ell}+ \frac{\delta^{im}_{\rm K}}{\bar{n}_i}\Big)   \Big( C^{jn}_{\ell}+\frac{\delta^{jn}_{\rm K}}{\bar{n}_j}\Big) + \Big(C^{in}_{\ell}+\frac{\delta^{in}_{\rm K}}{\bar{n}_i }\Big) \Big(C^{jm}_{\ell}+\frac{\delta^{jm}_{\rm K}}{\bar{n}_j}\Big) \Big], 
\end{align}
where $ \bar{\mathcal{ L}}_\ell $ stands for the bin-averaged Legendre polynomial, while $f_{\rm sky} $ represents the fraction of sky coverage, and $ \delta_{\rm K} $ denotes the Kronecker delta. The Poisson shot noise is considered, and $ \bar{n}_i $ denotes the angular number density in bin $i$. We use the {\sc camb sources} code \citep{Lewis:2007kz} to compute $C_\ell $. Among other things, we highlight that $ \bar{\mathcal{ L}}_\ell $ accounts for the finite binning width in configuration space  \citep{Salazar-Albornoz:2016psd}. Moreover, later in \citet{DESY6_BAO}, we further improve the modeling by including the mask correction effect \citep{cosmolike_mask}, which is important when the surface number density is low.  We have tested this covariance against DES mock catalog in \citet{Chan:2018gtc} and with the  Gaussian covariance from {\sc Cosmolike} \citep{cosmolike,cosmolike2020}, and both found good agreement.





\begin{figure}
    \centering
    \includegraphics[width=\linewidth]{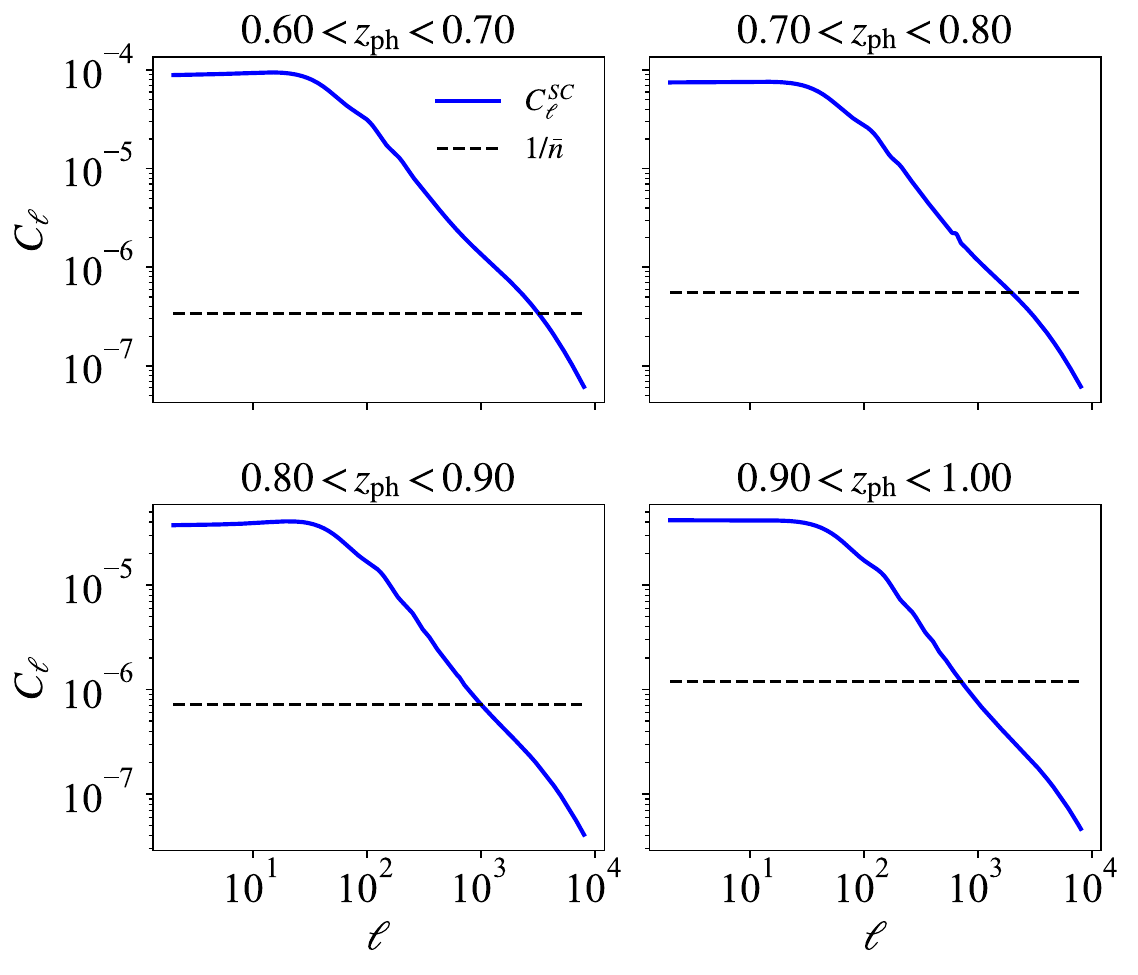}
    \caption{The $C_{\ell}$ (blue solid) and  shot noise (black dashed) for the tomographic bins.  Up to the BAO scale ($\ell \lesssim 400 $), the shot noise is subdominant compared to the signal $C_\ell $. 
    }
    \label{fig:Cl_shot_noise}
\end{figure}

In Fig.~\ref{fig:Cl_shot_noise}, we compare the $C_\ell $ against the Poisson shot noise. We plug these ingredients into Eq.~\eqref{eq:cov_mat_crossz} to compute the Gaussian covariance. The BAO scale extends up to  $ \ell \sim 400 $ at $z= 1$, and we see that the Poisson shot noise is subdominant up to this scale in all the tomographic bins.  

\begin{figure}
    \centering
    \includegraphics[height=2.5in]{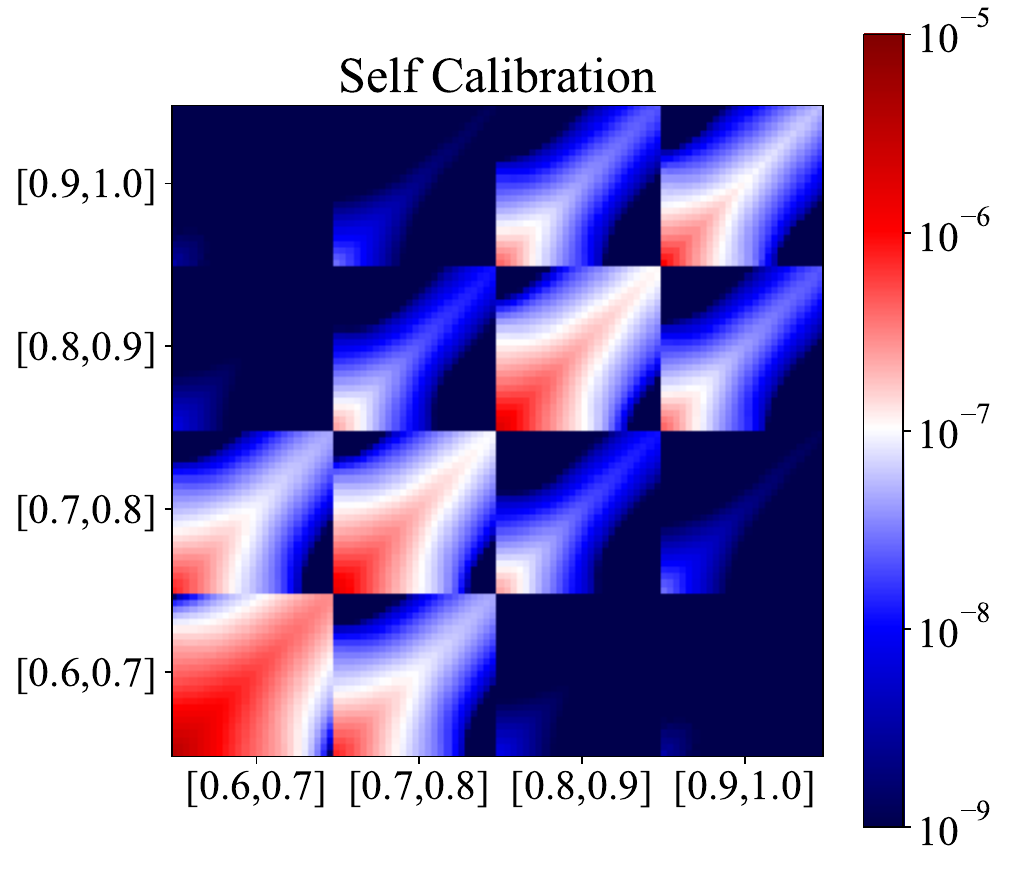}
    \caption{The covariance matrix of angular correlation functions of the four tomographic bins used for BAO fit. It is computed with the SC redshift distribution. Each sub-block is a $25\times25$ matrix.}
    \label{fig:SC_cov}
\end{figure}
 
Fig.~\ref{fig:SC_cov} shows the result of Gaussian covariance matrices for the four tomographic bins. In each sub-block, the $\theta$ values  ranges from 0 to
5 degree in angular interval of 0.2 degree, so that there are 25 angular bins in each sub-block. 
The Gaussian covariance is dominated by the diagonal and the correlation dies off slowly as the angle of separation increases. We also see some significant correlation between the neighboring tomographic bins. This is a manifestation that the measurements in configuration space is highly correlated, and a proper covariance is required to interpret the results.

We also contrast the Gaussian covariance results with the jackknife covariance ones. The jackknife covariance is computed similar to \citet{XU2023}, \change{where we divided the footprint into 120 spatially contiguous and equally sized subregions using a k-means clustering algorithm\footnote{\url{https://github.com/esheldon/kmeans_radec}}. When computing the covariance matrix,  we did not apply the correction in  \citet{Mohammad_Percival2022}. Our jackknife covariance may overestimate the true one by 25-60\% \citep{Norberg09}. However, we found that our results seem not too sensitive to the choice of the covariance matrix given that the best fit is shifted by 0.3$\sigma$ when switching from the Gaussian covariance matrix to the jackknife one. In addition, since we have not thoroughly validated the jackknife covariance for the BAO measurement, we only take it as a secondary choice
}. 

\subsection{ Parameter inference}
\label{sec:parameter_inference}

In this subsection, we describe the procedures to extract the BAO scale from the angular correlation measurement, \change{which  follows the angular correlation function analysis of DES Y3 \citep{DES:2021esc}. }  In the template fitting method, the template and covariance are evaluated in a fiducial cosmology, which is the Planck cosmology in this work.  To account for the difference in cosmology between the fiducial cosmology and data cosmology, the BAO scale dilation parameter $\alpha $ is introduced.  In term of the comoving angular diameter distance $ D_{\rm M} $ and sound horizon $r_{\rm s} $, $\alpha $ is defined as
\beq
\label{eq:alpha_DMrs}
\alpha = \frac{ D_{\rm M} r_{\rm s}^{\rm fid} }{ D_{\rm M}^{\rm fid}  r_{\rm s} }, 
\eeq
where the quantities with and without "fid" means that they are evaluated in the fiducial and the data cosmology, respectively.  

The full model is given by 
\begin{equation}
    w_{\text{model}}(\theta)=Bw_{\text{temp}}(\alpha\theta)+\sum\limits_{i}\frac{A_i}{\theta^i},
    \label{eq:w_model}
\end{equation}
where $w_{\text{temp}}$ denotes the theoretical template described in Sec.~\ref{sec:BAO_template} and $B$ is an amplitude parameter. The polynomial in $ 1/\theta $ is a smooth function introduced to absorb the imperfectness in the modeling of the broadband shape of the correlation due to e.g.~the nonlinear effect and the difference in cosmology between the fiducial cosmology and data cosmology. In the fiducial setup, $i$ goes over 0, 1, and 2.  While for each tomographic bin, there is a set of $B $ and $A_i $,  the shift of $\alpha$ is caused by the difference in cosmology model and so there is a single $\alpha$ for all the bins.

Under the assumption of Gaussian likelihood, the likelihood  $ \propto e^{- \chi^2 / 2 } $, we can extract the best fit parameter by minimizing the $\chi^2 $, which is given by
\begin{equation}
    \chi^2=\sum\limits_{i,j}[w_\text{obs}(\theta_i)-w_\text{model}(\theta_i)]C^{-1}_{ij}[w_\text{obs}(\theta_j)-w_\text{model}(\theta_j)],
\end{equation}
where $w_\text{obs}$ is the angular correlation function measurement   and $C$ is the covariance matrix.

Our maximum likelihood estimation procedure follows the steps in \citet{Chan:2018gtc}. We sample $\alpha$ over the range of [0.8,1.2] uniformly to get its $\chi^2 $ profile. To do so, for each  $\alpha $, we first analytically fit $A_i$, taking advantage of the fact that they are linear parameters. We then numerically fit $B$ with the constraint $B>0$. Finally, we end up with the $\chi^2 $ for $ \alpha $.  From the  $\chi^2 $ profile, we can derive the best fit and its 1-$\sigma$ error bar.  The best fit $\alpha $ locates at the position where the minimum $\chi^2 $,  $\chi^2_{\rm min} $, occurs. We adopt the error bar defined by half of the $\alpha$  region width, whose end points are given by the deviation of $\chi^2$  from $\chi^2_{\rm min} $ by 1. We define BAO to be detected if the whole 1  $\sigma$ interval enclosing the best fit falls within the interval [0.8,1.2].


\section{BAO fit results}
\label{sec:results}

After discussing the analysis pipeline, we are ready to apply it to our magnitude limited galaxy sample to extract the BAO signal. In this section, we first present our BAO fit results and then consider various tests to demonstrate the robustness of the measurement.


\begin{figure*}
    \centering
    \includegraphics[width=5.5in]{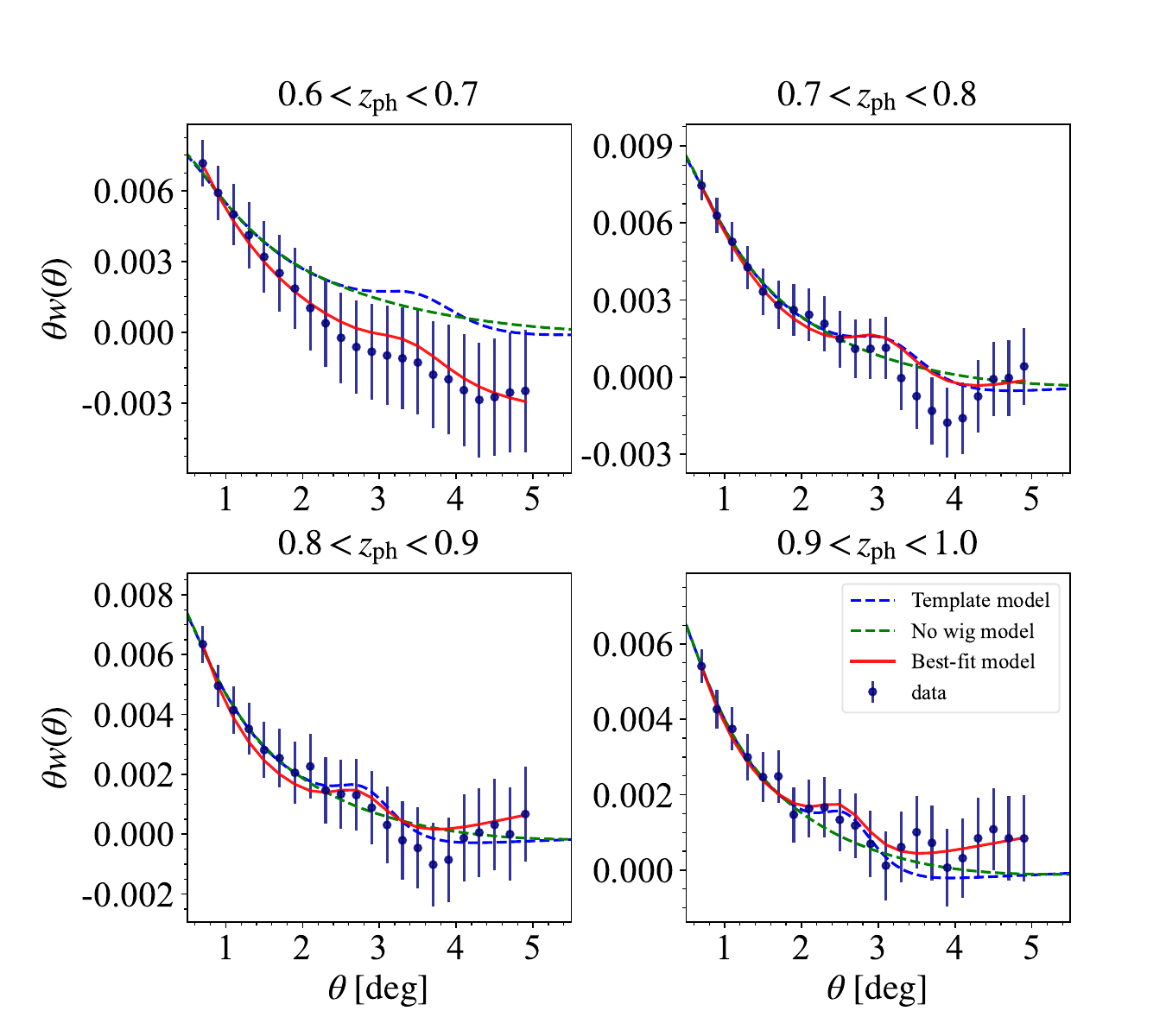}
    \caption{The BAO fit results obtained with the SC redshift distribution. The markers show the systematics mitigated measurements, and their error bars are derived from the Gaussian covariance. The blue dashed lines are the original templates generated from fiducial \change{Planck} cosmology, and the red lines are the full best fit model. In comparison, the   no-wiggle best fit results (green dashed) are also overplotted. 
    }
    \label{fig:BAO_fit_SC}
\end{figure*}

\subsection{ Main results }

In Fig.~\ref{fig:BAO_fit_SC}, we compare the angular correlation function measurements against the best fit model obtained using SC distribution. The BAO feature is visible around 3 degree. As redshift decreases, the angular BAO scale moves from $\sim 2.5 $ degree to  $\sim 3.5 $ degree.  The data measurements agree well with the best fit model, which gives  $\alpha=1.025\pm 0.033$ with $\chi^2 = 51.6$ for 71 degrees of freedom.

For fiducial  Planck cosmology, $r_{\mathrm{s}}=147.6  \, \mathrm{Mpc}$ and $D_{\mathrm{M}} (z_{\text {eff }} = 0.749  )=2726.7  \, \mathrm{Mpc}$.
Plugging these into Eq.~\eqref{eq:alpha_DMrs}, we find the ratio $ D_{\rm M} / r_{\rm s} $ constrained to be $18.94 \pm 0.61$.

Among the four tomographic bins, the BAO signal is lowest in the first bin because the  BAO damping is stronger at low $z$. The covariance at the same comoving scale is also larger at late time. Moreover, as redshift decreases,  the BAO angular scales move to a larger angular scale, and this reduces the number of pairs available. All these cause the signal-to-noise ratio of the first bin to be the lowest. For the first bin,  we also note that the shape of the original template does not match the data well, especially around the BAO region. This could be caused by over-correction by the mitigation procedure. But we will see that this does not really affect the BAO measurement. We have also plotted the best fit results obtained with the no-wiggle template. Visually, the BAO template results in a better fit than the no-wiggle one.  In contrast the weighted spec-$z$ distribution results in $\alpha=1.018\pm 0.030$, in good agreement with the SC results.

\begin{figure}
    \centering
    \includegraphics[width=3in]{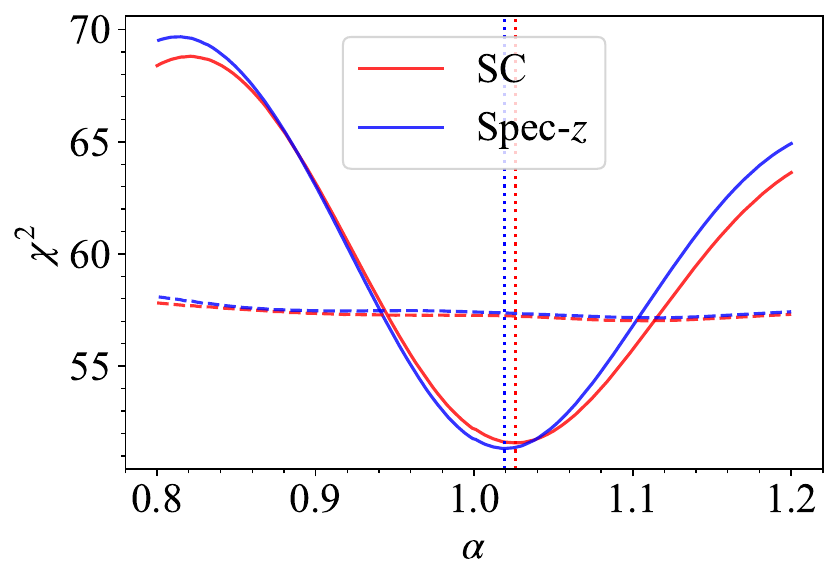}
    \caption{The  $\chi^2$ profiles  obtained from the SC (red) and weighted spec-$z$ (blue) template. The dash lines show the corresponding no-wiggle template results. The dotted lines indicate the best fit $\alpha$. \change{The fiducial cosmology is Planck, corresponding to $\alpha =1 $.  } 
    }
    \label{fig:chi2_SC}
\end{figure}

We compare the $\chi^2$ profile  obtained with the SC and weighted spec-$z$ distribution in Fig.~\ref{fig:chi2_SC}. The profiles are consistent with each other. In addition, we also plot the profiles obtained with the no-wiggle templates, which are essentially flat in the whole range. Since the $\chi^2$ is well approximated by a quadratic, the likelihood function for $\alpha $ is approximately Gaussian, we can use this probability function to quantify the significance of BAO detection. Because $ \chi^2$ difference between the wiggle and no-wiggle fit is about 6, this translates our BAO detection significance to 2.4 $\sigma$.

\subsection{Robustness test}
\label{sec:RobustnessTest}

To check the robustness of our measurements, we conduct a number of tests with different analysis setup and the results are shown in Table ~\ref{tab:robustness}. 
Here we discuss the details of the tests. 
\begin{itemize}

\item  The magnification bias plays an important role in the SC $n(z)$ estimation.   We first test its impact by considering the $n(z)$ from the SC method using the correlation measurement  without magnification bias correction applied at all. We find that the best fit $\alpha$ is reduced by half $\sigma$ in this case. This demonstrates that the magnification bias correction must be applied otherwise a sizable bias will be induced. \change{ Next we consider variations of the magnification bias correction treatment.  Recall that to compute the cosmic magnification correction, we need the true-$z$ distribution of two separate tomographic bins, the linear bias of the near bin, and the slope of the luminosity function of the distant bin at the limiting magnitude.  We consider $s$/2 and $2s$, where $s$ is the fiducial slope at the limiting flux. In this case, the maximum change occurs ($ 0.4 \sigma $) when $s$/2 is used. However, we note that our variations are substantial, as suggested by \citet{vonWietersheim-Kramsta2021}. We also consider the galaxy bias parameters obtained using the  SC or weighted spec-$z$  $n(z)$ following the treatment mentioned at the end of Sec.~\ref{sec:BAO_template}.  We find that the resultant $\alpha$ changes by  $ 0.09 \sigma $ and $0.24 \sigma $, respectively. These show that the results have some sensitivity on the galaxy bias of the sample. When the weighted spec-$z$ $n(z)$ is used instead, the result is unchanged.     }

\item As we mentioned, the weighted spec-$z$ result is similar to the SC one, with the shift in best fit by about 0.2 $\sigma$. \change{ To test the impact of coarse binning in $n(z)$, we show the results obtained by downsampling the fiducial fine spec-$z$ $n(z)$ to the resolution of the SC $n(z)$ ($\Delta z = 0.1$). There is little change in the results thanks to the cubic spline applied. }   \change{ In these tests, we change only the template while leaving the covariance computed in the fiducial SC $n(z)$. We go on to show the case when both the template and covariance are computed in weighted spec-$z$ $n(z) $. We find that only the error bar changes slightly relatively to the previous case, while the best fit value remains unchanged.  }

\item We have applied the weight to mitigate the systematics due to survey properties. To check its importance for BAO fit, we have performed the BAO fit on the correlation measurement without systematics treatment. We find that the fit results essentially do not depend on the mitigation weight, in agreement with \citet{Abbott:2017wcz, DES:2021esc}. This is because BAO is a sharp feature, while the systematics generally give rise to smooth contributions. 

\item The fiducial covariance is the theory Gaussian covariance.  We consider the alternative jackknife covariance, which is computed as in \citet{XU2023}. The best fit is shifted by 0.3 $\sigma $, but the error bar remains almost the same.  We only take this test to be informative as the jackknife covariance has not been thoroughly validated for BAO measurement.

\item The default template is computed in Planck cosmology. To check the dependence on the template cosmology, we consider the template computed in MICE cosmology, as in \citet{Abbott:2017wcz, DES:2021esc}. MICE cosmology \citep{Fosalba_etal2015,Crocce_etal2015} is a flat $ \Lambda$CDM model  with $\Omega_{\rm m} = 0.25$,  $\Omega_{\Lambda} = 0.75$, $h=0.7$, and $\sigma_8 = 0.8$.   In MICE cosmology, $r_{\mathrm{s}}=153.4 \, \mathrm{Mpc}$ and $D_{\mathrm{M}}\left(z_{\text {eff }}\right) = 2712.9 \,  \mathrm{Mpc}$. Therefore, we expect the best fit $\alpha$ obtained in MICE template to be larger than the Planck result by a factor of 1.045.   By taking this factor into account, the nominal best fit $1.071 \pm 0.036 $ is transformed to $1.025 \pm 0.034 $, which is in nice agreement with the Planck template case.

\item   The default angular range for BAO fit is [0.5,5] degree. We test reducing the minimal $\theta $ to 1 degree or the maximum degree to 4 degree. The results are consistent with the fiducial results, with little changes.

\item The default bin width is 0.2 degree. We consider an alternative binning width of 0.05 degree. In this case, although the $\chi^2 / {\rm dof} $ is inflated to 1.25, the best fit $1.023\pm0.035 $ remains consistent with the fiducial one.  

\end{itemize}

\begin{figure}
    \centering
    \includegraphics[width=3.2in]{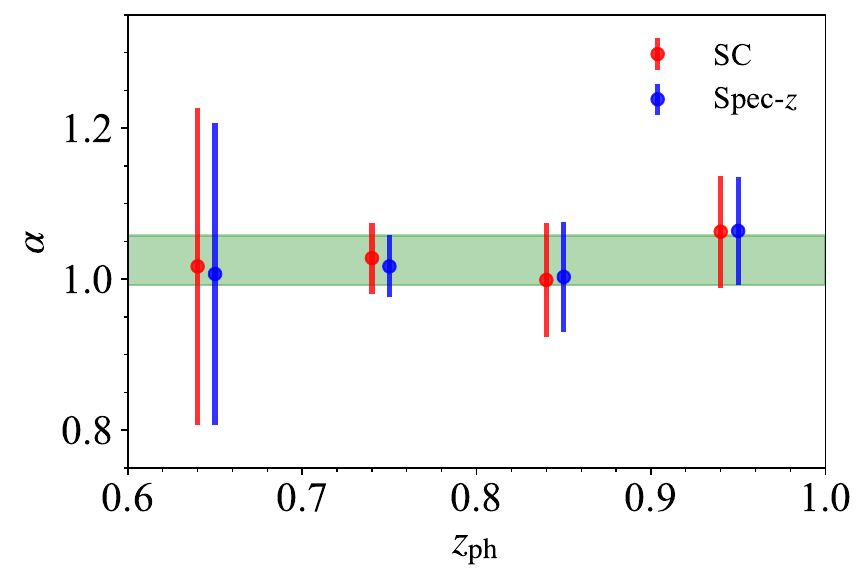}
    \caption{Best fit $\alpha$ for individual tomographic bins for SC (red) and weighted spec-$z$ (blue) results. The green band shows the 1-$\sigma$ range for our fiducial model. The error bar size in the first bin is so wide that it exceeds 0.2 and hence BAO is not detected according to our detection criterion.  We simply use an arbitrary value to cover the range from [0.8,1.2] instead. There are slight horizontal offsets to avoid clutter. }
    \label{fig:individual_bin_comp}
\end{figure}

In the lower part of Table ~\ref{tab:robustness}, we show the fit results to the individual tomographic bins for both the SC and weighted spec-$z$ results. Alternatively, we plot these individual bin results together with the fiducial four-bin joint result in Fig.~\ref{fig:individual_bin_comp}.  The signal-to-noise ratio of the first bin is so weak that although its best fit is quite central, the error bar derived is so wide that it extends out of the interval [0.8,1.2]. According to our BAO detection criterion given in Sec.~\ref{sec:parameter_inference}, we do not detect BAO in the first bin. This is true for both SC and weighted spec-$z$ results.  Thus we only show the best fit values in the table and assign the error size of 0.2  arbitrarily in Fig.~\ref{fig:individual_bin_comp}.  We detect BAO in the rest of the tomographic bins, and the signals are consistent with the fiducial 1 $\sigma$ bound.  \change{The largest difference occurs in the second bin ($ \sim 0.3 \sigma $), and this echoes with the relatively large difference between the SC and weighted spec-$z$ $n(z) $ observed in Fig.~\ref{fig:SC_z_distribution}.     }

\begin{table}
\centering
\caption{ Various robustness test results. For each test, the best fit and the associated $\chi_{\rm min}^2 / {\rm dof}$  are shown.  In the lower parts of the table, the fit results for individual bins obtained with the SC and weighted spec-$z$ templates are presented.   }
\begin{tabular}{lcc}
  \hline
  \hline
   \bf{Case}           & $\alpha$         &  $\chi_{\rm{min}}^2$/dof \\
  \hline
  \textbf{ Fiducial}     & $1.025\pm 0.033$     & $51.6/71 (=0.73)$         \\
 \hline
  \change{\textbf{Magnification bias corr.}} \\
    No correction & $1.008\pm0.032$           &    $52.0/71(=0.73)$                    \\
\change{  $s \times0.5$ } & $1.011\pm0.031 $ & $51.7/71(=0.73)$\\
\change{  $s \times2$ } & $1.018\pm 0.031$ & $51.5/71(=0.72)$\\
\change{ SC bias } & $1.028\pm 0.033$ & $51.6/71(=0.73)$\\
\change{ Spec-$z$ bias} & $1.017\pm 0.033$ & $51.5/71(=0.73)$\\
\change{ Spec-$z$   $n(z)$ }& $1.025\pm 0.033$ & $51.7/71(=0.73)$\\   
  \hline
 Spec-$z$ $n(z)$ &    $1.018\pm 0.030$   & $51.3/71(=0.72)$  \\
\change{Coarse spec-$z$ $n(z)$}& $1.017\pm 0.032$& $51.6/71(=0.73)$ \\
\change{Spec-$z$ $n(z)$ template+cov}&    $1.018\pm 0.033$   & $56.2/71(=0.79)$  \\
  No $w_{\rm sys}$ &    $1.025\pm 0.034$   & $52.8/71=(0.74)$  \\
  Jackknife cov &    $1.016\pm  0.031$   & $52.8/71(=0.74)$ \\
   MICE template \change{/ 1.045}  &    \change{$1.025\pm 0.034$}   & $51.8/71(=0.73)$  \\
   $\theta_{\rm min} = 1^{\circ}$  & $1.024\pm 0.033$ & $49.1/63(=0.78)$ \\
   $ \theta_{\rm max} = 4^{\circ}$  & $1.025\pm0.036$ &$39.7/51=(0.78)$ \\
  $\Delta \theta =0.05^{\circ}$ & $1.023\pm 0.035$ & $430.0/343(=1.25)$  \\
 \hline
  \textbf{Individual bin: SC}    \\
  $0.6<z_\text{ph}<0.7$ & $1.017 $ & $2.2/17(=0.13)$\\
  $0.7<z_\text{ph}<0.8$ & $1.028\pm 0.047$ & $14.2/17(=0.83)$\\
  $0.8<z_\text{ph}<0.9$ & $0.999\pm 0.075$ & $16.4/17(=0.97)$\\
  $0.9<z_\text{ph}<1.0$ & $1.063\pm 0.074$ & $18.7/17(=1.10)$\\
    \hline
   \textbf{Individual bin: spec-$z$} \\
  $0.6<z_\text{ph}<0.7$ & $1.007 $ & $2.2/17(=0.13)$\\
  $0.7<z_\text{ph}<0.8$ & $1.017\pm 0.041$ & $13.8/17(=0.81)$\\
  $0.8<z_\text{ph}<0.9$ & $1.003\pm 0.073$ & $16.4/17(=0.97)$\\
  $0.9<z_\text{ph}<1.0$ & $1.064\pm 0.072$ & $18.8/17(=1.10)$\\
    \hline    
   \hline
  \label{tab:robustness}
\end{tabular}
\end{table}

\subsection{ Comparison with DES BAO}

Here we would like to compare our results with the DES Y3 BAO measurement (hereafter Y3)\citep{DES:2021esc}, which was the most precise BAO measurement obtained from photometric data \change{until recently, which is only broken by itself, DES Y6 \citep{DESY6_BAO}. We will mainly compare with Y3 because its depth is closer to ours, but we will also highlight its main difference with Y6. }  Our analysis pipeline is very similar to Y3, thus our results differ mainly in the data sample used.

We only use DECaLS-NGC region, and so our dataset is distinct from the Y3 sample \citep{CarneroRosell_etal2022}. Y3 also have four tomographic bins coincide with ours, so that we can directly compare our results with theirs bin-wisely. Y3 footprint covers an area of 4108 deg$^2$, comparable to ours. While we simply apply the z-band cut, $z_{\rm mag}< 21 $, they apply color cuts to select a red galaxy sample \change{ and further optimize it to enhance the BAO constraint} \citep{Crocce_etal2019}.   This allows us to have more galaxies bin-wisely relative to Y3. For example, the total number of galaxies is 10.4 million for us and 6.1 million for Y3. The bias parameters of Y3 are 1.8, 1.8, 2.0 and 2.1 for the four tomographic bins.  They are generally higher than ours because of the red galaxy nature of their sample. Overall, the photo-$z$ accuracy of the DES sample is slightly better, with the spread of  the true-$z$ distribution smaller by  $\sim  13 \%$. Using only the four bins coincident with ours, the Y3 constraint on $\alpha$  is $0.967\pm 0.026$ with respect to the Planck cosmology. \change{ The best fit is about 2 $\sigma$  from ours, and its error bar is 79 \% of ours. }
Even though our magnitude limited sample has more galaxies, higher bias parameters and better photo-$z$ quality are more vital in achieving a tighter BAO constraint.

\change{ The Y6 sample \citep{DESY6_BAOsample} occupies a similar footprint as Y3, and in fact its many properties are similar to the Y3 one. The main difference is that the Y6 sample is much deeper in magnitude and one of the consequences is that the total number of galaxies in the range [0.6,1.0] reaches 13.6 million. Its photo-$z$ quality is also higher, with the spread being reduced by $\sim 5\%$ relative to Y3. Their $ w(\theta) $ analysis in the redshift range [0.6,1.0] yields $\alpha =0.945 \pm 0.024 $ with respect to Planck. This is consistent with the Y3 results, but it increases the discrepancy from ours to more than 2 $\sigma$. This intriguing difference could arise from sheer statistical fluctuation or unknown systematics, but, more interestingly, it may hint some new physics. } 

In principle, our constraint can be further improved by also using the south cap region, \change{and this can also crosschecks the DES results in the same sky region}. However, as our goal here is to mainly test the usage of the SC distribution for BAO measurement, we shall not pursue this here.

\section{Conclusions}
\label{sec:conclusions}

With the advancement of the imaging surveys, the photometric BAO has been measured with high precision recently. To get an unbiased BAO measurement, the true-$z$ distribution of the photo-$z$ data must be properly calibrated. In this work we consider the calibration using the SC method. Previous works on the applications of the SC are more oriented towards weak lensing, and this is the first one aiming for BAO measurement.

We use a magnitude limited sample with  $z_{\rm mag} < 21 $ derived from PRLS, which is further based on LS DR9.  The sample consists of 10,576,969 galaxies in an area of 4974 deg$^2$. The redshift range of the sample is from 0.6 to 1, and we divide it into four photo-$z$ bins, each of bin width $\Delta z = 0.1 $. We use the SC method to calibrate the true-$z$ distribution of the photo-$z$ bins. This method relies on the premise that the angular correlation between different photo-$z$ bins is only caused by the physically overlapping galaxies. Our SC distribution is obtained by solving the scattering probability in full generality.  It also includes the magnification bias correction, which is an important source of clustering when the bins are widely separated. However, to avoid highly correlated measurement, the bin width cannot be set too fine.  This limits the redshift resolution of the true-$z$ distribution. For the weighted spec-$z$ distribution, if the number of spec-$z$ galaxies are sufficient, its resolution can be set to a finer scale.


We measure the BAO using the angular correlation function.   Our fiducial constraint on the BAO scale dilation parameter is  $\alpha = 1.025\pm 0.033$, or  the ratio $ D_{\rm M} / r_{\rm s} = 18.94 \pm 0.61$ at the effective redshift of the sample $z_{\rm eff} =0.749 $.  The results are similar to the weighted spec-$z$ ones, which yield  $\alpha = 1.018 \pm 0.030 $.  In particular, this demonstrates that  the resolution problem does not seriously affect our results.  Furthermore, we perform a series of robustness tests on the BAO fit results, and find that the variation is within 1$\sigma $. \change{ Interestingly, our measurements are well consistent with the fiducial Planck cosmology, and hence it is discrepant with the latest DES Y6 measurement \citep{DESY6_BAO} at a similar effective redshift but different part of the sky, by more than 2 $\sigma $. }

Our results show that SC methods can be used to calibrate the distribution for the BAO measurements. This method is expected to play a more important role as we explore the high redshift universe, where the spec-$z$ samples are much rarer, and hence the utility of the weighted spec-$z$ and clustering-$z$ methods are limited.    \change{Furthermore, the method can be strengthened by combining the information of the SC with the clustering-$z $ when there is some external overlapping spec-$z$ data available. }   We plan to further explore these applications, especially in higher redshifts in the future.

\section*{Acknowledgments}

\change{ We thank Eoin Colg\'ain for useful discussions. We are also grateful to the anonymous referee for his/her constructive comments that improve the presentation of the manuscript. }. This work  is supported by the National Science Foundation of China under the grant number 12273121 and the science research grants from the China Manned Space Project. HX is supported by the National SKA Program of China (grant No. 2020SKA0110100), the National Natural Science Foundation of China (Nos. 11922305, 11833005) and the science research grants from the China Manned Space Project with NOs. CMS-CSST-2021-A02. We acknowledge the use of the Gravity Supercomputer at the Department of Astronomy, Shanghai Jiao Tong University.
 
 The Photometric Redshifts for the Legacy Surveys (PRLS) catalog used in this paper was produced thanks to funding from the U.S. Department of Energy Office of Science,  Office of High Energy Physics via grant DE-SC0007914.
{The DESI Legacy Imaging Surveys consist of three individual and complementary projects: the Dark Energy Camera Legacy Survey (DECaLS), the Beijing-Arizona Sky Survey (BASS), and the Mayall z-band Legacy Survey (MzLS). DECaLS, BASS and MzLS together include data obtained, respectively, at the Blanco telescope, Cerro Tololo Inter-American Observatory, NSF’s NOIRLab; the Bok telescope, Steward Observatory, University of Arizona; and the Mayall telescope, Kitt Peak National Observatory, NOIRLab. NOIRLab is operated by the Association of Universities for Research in Astronomy (AURA) under a cooperative agreement with the National Science Foundation. Pipeline processing and analyses of the data were supported by NOIRLab and the Lawrence Berkeley National Laboratory. Legacy Surveys also uses data products from the Near-Earth Object Wide-field Infrared Survey Explorer (NEOWISE), a project of the Jet Propulsion Laboratory/California Institute of Technology, funded by the National Aeronautics and Space Administration. Legacy Surveys was supported by: the Director, Office of Science, Office of High Energy Physics of the U.S. Department of Energy; the National Energy Research Scientific Computing Center, a DOE Office of Science User Facility; the U.S. National Science Foundation, Division of Astronomical Sciences; the National Astronomical Observatories of China, the Chinese Academy of Sciences and the Chinese National Natural Science Foundation. LBNL is managed by the Regents of the University of California under contract to the U.S. Department of Energy. The complete acknowledgments can be found at https://www.legacysurvey.org/.
Any opinions, findings, and conclusions or recommendations expressed in this material are those of the author(s) and do not necessarily reflect the views of the U. S. National Science Foundation, the U. S. Department of Energy, or any of the listed funding agencies.
The DESI collaboration are honored to be permitted to conduct scientific research on Iolkam Du’ag (Kitt Peak), a mountain with particular significance to the Tohono O’odham Nation.}

\section*{Data Availability}
The data underlying this article will be shared on reasonable request to the corresponding authors.



\bibliographystyle{mnras}
\bibliography{references}

\begin{thebibliography}{}
\makeatletter
\relax
\def\mn@urlcharsother{\let\do\@makeother \do\$\do\&\do\#\do\^\do\_\do\%\do\~}
\def\mn@doi{\begingroup\mn@urlcharsother \@ifnextchar [ {\mn@doi@}
  {\mn@doi@[]}}
\def\mn@doi@[#1]#2{\def\@tempa{#1}\ifx\@tempa\@empty \href
  {http://dx.doi.org/#2} {doi:#2}\else \href {http://dx.doi.org/#2} {#1}\fi
  \endgroup}
\def\mn@eprint#1#2{\mn@eprint@#1:#2::\@nil}
\def\mn@eprint@arXiv#1{\href {http://arxiv.org/abs/#1} {{\tt arXiv:#1}}}
\def\mn@eprint@dblp#1{\href {http://dblp.uni-trier.de/rec/bibtex/#1.xml}
  {dblp:#1}}
\def\mn@eprint@#1:#2:#3:#4\@nil{\def\@tempa {#1}\def\@tempb {#2}\def\@tempc
  {#3}\ifx \@tempc \@empty \let \@tempc \@tempb \let \@tempb \@tempa \fi \ifx
  \@tempb \@empty \def\@tempb {arXiv}\fi \@ifundefined
  {mn@eprint@\@tempb}{\@tempb:\@tempc}{\expandafter \expandafter \csname
  mn@eprint@\@tempb\endcsname \expandafter{\@tempc}}}

\bibitem[\protect\citeauthoryear{Abbott et~al.}{Abbott
  et~al.}{2019}]{Abbott:2017wcz}
Abbott T.,  et~al., 2019, \mn@doi [Mon. Not. Roy. Astron. Soc.]
  {10.1093/mnras/sty3351}, 483, 4866

\bibitem[\protect\citeauthoryear{Abbott et~al.}{Abbott
  et~al.}{2022}]{DES:2021esc}
Abbott T. M.~C.,  et~al., 2022, \mn@doi [Phys. Rev. D]
  {10.1103/PhysRevD.105.043512}, 105, 043512

\bibitem[\protect\citeauthoryear{{Abbott} et~al.,}{{Abbott}
  et~al.}{2024}]{DESY6_BAO}
{Abbott} T.~M.~C.,  et~al., 2024, \mn@doi [arXiv e-prints]
  {10.48550/arXiv.2402.10696}, \href
  {https://ui.adsabs.harvard.edu/abs/2024arXiv240210696D} {p. arXiv:2402.10696}

\bibitem[\protect\citeauthoryear{{Alam}, {Ata}, {Bailey}, {Beutler}, {Bizyaev}
  et~al.}{{Alam} et~al.}{2017}]{Alam_etal2017}
{Alam} S.,  {Ata} M.,  {Bailey} S.,  {Beutler} F.,  {Bizyaev} D.,   et~al.,
  2017, \mn@doi [MNRAS] {10.1093/mnras/stx721}, \href
  {http://adsabs.harvard.edu/abs/2017MNRAS.470.2617A} {470, 2617}

\bibitem[\protect\citeauthoryear{Alam et~al.}{Alam
  et~al.}{2021}]{eBOSS:2020yzd}
Alam S.,  et~al., 2021, \mn@doi [Phys. Rev. D] {10.1103/PhysRevD.103.083533},
  103, 083533

\bibitem[\protect\citeauthoryear{{Amendola}, {Quercellini}  \&
  {Giallongo}}{{Amendola} et~al.}{2005}]{Amendola_etal2005}
{Amendola} L.,  {Quercellini} C.,   {Giallongo} E.,  2005, \mn@doi [MNRAS]
  {10.1111/j.1365-2966.2004.08558.x}, \href
  {https://ui.adsabs.harvard.edu/abs/2005MNRAS.357..429A} {357, 429}

\bibitem[\protect\citeauthoryear{{Anderson} et~al.}{{Anderson}
  et~al.}{2012}]{Anderson_BOSS2012}
{Anderson} L.,  et~al., 2012, \mn@doi [MNRAS]
  {10.1111/j.1365-2966.2012.22066.x}, \href
  {https://ui.adsabs.harvard.edu/abs/2012MNRAS.427.3435A} {427, 3435}

\bibitem[\protect\citeauthoryear{{Arnouts}, {Cristiani}, {Moscardini},
  {Matarrese}, {Lucchin}, {Fontana}  \& {Giallongo}}{{Arnouts}
  et~al.}{1999}]{Arnouts_1999}
{Arnouts} S.,  {Cristiani} S.,  {Moscardini} L.,  {Matarrese} S.,  {Lucchin}
  F.,  {Fontana} A.,   {Giallongo} E.,  1999, \mn@doi [\mnras]
  {10.1046/j.1365-8711.1999.02978.x}, \href
  {https://ui.adsabs.harvard.edu/abs/1999MNRAS.310..540A} {310, 540}

\bibitem[\protect\citeauthoryear{{Aubourg} et~al.}{{Aubourg}
  et~al.}{2015}]{Aubourg:2014yra}
{Aubourg} {\'E}.,  et~al., 2015, \mn@doi [Phys. Rev.]
  {10.1103/PhysRevD.92.123516}, D92, 123516

\bibitem[\protect\citeauthoryear{{Bartelmann} \& {Schneider}}{{Bartelmann} \&
  {Schneider}}{2001}]{BartelmannSchneider2001}
{Bartelmann} M.,  {Schneider} P.,  2001, \mn@doi [\physrep]
  {10.1016/S0370-1573(00)00082-X}, \href
  {https://ui.adsabs.harvard.edu/abs/2001PhR...340..291B} {340, 291}

\bibitem[\protect\citeauthoryear{{Ben{\'\i}tez}}{{Ben{\'\i}tez}}{2000}]{Benitez_2000}
{Ben{\'\i}tez} N.,  2000, \mn@doi [\apj] {10.1086/308947}, \href
  {https://ui.adsabs.harvard.edu/abs/2000ApJ...536..571B} {536, 571}

\bibitem[\protect\citeauthoryear{Benitez et~al.}{Benitez
  et~al.}{2009}]{Benitez:2008fs}
Benitez N.,  et~al., 2009, \mn@doi [Astrophys. J.]
  {10.1088/0004-637X/691/1/241}, 691, 241

\bibitem[\protect\citeauthoryear{{Benjamin}, {van Waerbeke}, {M{\'e}nard}  \&
  {Kilbinger}}{{Benjamin} et~al.}{2010}]{Benjamine_etal2010}
{Benjamin} J.,  {van Waerbeke} L.,  {M{\'e}nard} B.,   {Kilbinger} M.,  2010,
  \mn@doi [\mnras] {10.1111/j.1365-2966.2010.17191.x}, \href
  {https://ui.adsabs.harvard.edu/abs/2010MNRAS.408.1168B} {408, 1168}

\bibitem[\protect\citeauthoryear{{Benjamin} et~al.,}{{Benjamin}
  et~al.}{2013}]{Benjamine_etal2013}
{Benjamin} J.,  et~al., 2013, \mn@doi [\mnras] {10.1093/mnras/stt276}, \href
  {https://ui.adsabs.harvard.edu/abs/2013MNRAS.431.1547B} {431, 1547}

\bibitem[\protect\citeauthoryear{{Beutler}, {Blake}, {Colless}, {Jones}
  et~al.}{{Beutler} et~al.}{2011}]{Beutler_etal2011}
{Beutler} F.,  {Blake} C.,  {Colless} M.,  {Jones} D.~H.,   et~al., 2011,
  \mn@doi [MNRAS] {10.1111/j.1365-2966.2011.19250.x}, \href
  {http://adsabs.harvard.edu/abs/2011MNRAS.416.3017B} {416, 3017}

\bibitem[\protect\citeauthoryear{{Blake} \& {Bridle}}{{Blake} \&
  {Bridle}}{2005}]{BlakeBridle_2005}
{Blake} C.,  {Bridle} S.,  2005, \mn@doi [\mnras]
  {10.1111/j.1365-2966.2005.09526.x}, \href
  {https://ui.adsabs.harvard.edu/abs/2005MNRAS.363.1329B} {363, 1329}

\bibitem[\protect\citeauthoryear{{Blake} et~al.}{{Blake}
  et~al.}{2012}]{Blake2012_WiggleZ}
{Blake} C.,  et~al., 2012, \mn@doi [MNRAS] {10.1111/j.1365-2966.2012.21473.x},
  \href {https://ui.adsabs.harvard.edu/abs/2012MNRAS.425..405B} {425, 405}

\bibitem[\protect\citeauthoryear{{Bolzonella}, {Miralles}  \&
  {Pell{\'o}}}{{Bolzonella} et~al.}{2000}]{Bolzonella_etal2000}
{Bolzonella} M.,  {Miralles} J.~M.,   {Pell{\'o}} R.,  2000, \mn@doi [\aap]
  {10.48550/arXiv.astro-ph/0003380}, \href
  {https://ui.adsabs.harvard.edu/abs/2000A&A...363..476B} {363, 476}

\bibitem[\protect\citeauthoryear{{Bond} \& {Efstathiou}}{{Bond} \&
  {Efstathiou}}{1984}]{BondEfstathiou1984}
{Bond} J.~R.,  {Efstathiou} G.,  1984, \mn@doi [ApJL] {10.1086/184362}, \href
  {http://adsabs.harvard.edu/abs/1984ApJ...285L..45B} {285, L45}

\bibitem[\protect\citeauthoryear{{Bond} \& {Efstathiou}}{{Bond} \&
  {Efstathiou}}{1987}]{BondEfstathiou1987}
{Bond} J.~R.,  {Efstathiou} G.,  1987, \mn@doi [MNRAS]
  {10.1093/mnras/226.3.655}, \href
  {http://adsabs.harvard.edu/abs/1987MNRAS.226..655B} {226, 655}

\bibitem[\protect\citeauthoryear{{Bonnett} et~al.,}{{Bonnett}
  et~al.}{2016}]{Bonnett_etal2016}
{Bonnett} C.,  et~al., 2016, \mn@doi [\prd] {10.1103/PhysRevD.94.042005}, \href
  {https://ui.adsabs.harvard.edu/abs/2016PhRvD..94d2005B} {94, 042005}

\bibitem[\protect\citeauthoryear{{Camacho} et~al.,}{{Camacho}
  et~al.}{2019}]{Camacho_DESY1Cl}
{Camacho} H.,  et~al., 2019, \mn@doi [\mnras] {10.1093/mnras/stz1514}, \href
  {https://ui.adsabs.harvard.edu/abs/2019MNRAS.487.3870C} {487, 3870}

\bibitem[\protect\citeauthoryear{{Carnero Rosell}, {Rodriguez-Monroy},
  {Crocce}, {Elvin-Poole}, {Porredon}  et~al.}{{Carnero Rosell}
  et~al.}{2022}]{CarneroRosell_etal2022}
{Carnero Rosell} A.,  {Rodriguez-Monroy} M.,  {Crocce} M.,  {Elvin-Poole} J.,
  {Porredon} A.,   et~al., 2022, \mn@doi [\mnras] {10.1093/mnras/stab2995},
  \href {https://ui.adsabs.harvard.edu/abs/2022MNRAS.509..778C} {509, 778}

\bibitem[\protect\citeauthoryear{{Carnero}, {S{\'a}nchez}, {Crocce},
  {Cabr{\'e}}  \& {Gazta{\~n}aga}}{{Carnero} et~al.}{2012}]{Carnero_etal2012}
{Carnero} A.,  {S{\'a}nchez} E.,  {Crocce} M.,  {Cabr{\'e}} A.,
  {Gazta{\~n}aga} E.,  2012, \mn@doi [MNRAS]
  {10.1111/j.1365-2966.2011.19832.x}, \href
  {http://adsabs.harvard.edu/abs/2012MNRAS.419.1689C} {419, 1689}

\bibitem[\protect\citeauthoryear{{Cawthon} et~al.}{{Cawthon}
  et~al.}{2022}]{Cawthon_etal2022}
{Cawthon} R.,  et~al., 2022, \mn@doi [\mnras] {10.1093/mnras/stac1160}, \href
  {https://ui.adsabs.harvard.edu/abs/2022MNRAS.513.5517C} {513, 5517}

\bibitem[\protect\citeauthoryear{Chan et~al.}{Chan et~al.}{2018}]{Chan:2018gtc}
Chan K.~C.,  et~al., 2018, \mn@doi [Mon. Not. Roy. Astron. Soc.]
  {10.1093/mnras/sty2036}, 480, 3031

\bibitem[\protect\citeauthoryear{{Chan}, {Avila}, {Carnero Rosell}, {Ferrero},
  {Elvin-Poole}  et~al.}{{Chan} et~al.}{2022a}]{Chan_xip2022}
{Chan} K.~C.,  {Avila} S.,  {Carnero Rosell} A.,  {Ferrero} I.,  {Elvin-Poole}
  et~al., 2022a, \mn@doi [\prd] {10.1103/PhysRevD.106.123502}, \href
  {https://ui.adsabs.harvard.edu/abs/2022PhRvD.106l3502C} {106, 123502}

\bibitem[\protect\citeauthoryear{{Chan}, {Ferrero}, {Avila}, {Ross}, {Crocce}
  \& {Gazta{\~n}aga}}{{Chan} et~al.}{2022b}]{Chan_xiptheory2022}
{Chan} K.~C.,  {Ferrero} I.,  {Avila} S.,  {Ross} A.~J.,  {Crocce} M.,
  {Gazta{\~n}aga} E.,  2022b, \mn@doi [\mnras] {10.1093/mnras/stac340}, \href
  {https://ui.adsabs.harvard.edu/abs/2022MNRAS.511.3965C} {511, 3965}

\bibitem[\protect\citeauthoryear{{Chan}, {Lu}  \& {Wang}}{{Chan}
  et~al.}{2023}]{Chan_etal2023}
{Chan} K.~C.,  {Lu} G.,   {Wang} X.,  2023, \mn@doi [arXiv e-prints]
  {10.48550/arXiv.2311.12611}, \href
  {https://ui.adsabs.harvard.edu/abs/2023arXiv231112611C} {p. arXiv:2311.12611}

\bibitem[\protect\citeauthoryear{{Chaussidon} et~al.,}{{Chaussidon}
  et~al.}{2022}]{Chaussidon_etal2022}
{Chaussidon} E.,  et~al., 2022, \mn@doi [\mnras] {10.1093/mnras/stab3252},
  \href {https://ui.adsabs.harvard.edu/abs/2022MNRAS.509.3904C} {509, 3904}

\bibitem[\protect\citeauthoryear{Chaves-Montero, Angulo  \&
  Hern\'andez-Monteagudo}{Chaves-Montero et~al.}{2018}]{Chaves-Montero:2016nmw}
Chaves-Montero J.,  Angulo R.~E.,   Hern\'andez-Monteagudo C.,  2018, \mn@doi
  [Mon. Not. Roy. Astron. Soc.] {10.1093/mnras/sty924}, 477, 3892

\bibitem[\protect\citeauthoryear{{Chisari} et~al.,}{{Chisari}
  et~al.}{2019}]{cclpaper}
{Chisari} N.~E.,  et~al., 2019, \mn@doi [\apjs] {10.3847/1538-4365/ab1658},
  \href {https://ui.adsabs.harvard.edu/abs/2019ApJS..242....2C} {242, 2}

\bibitem[\protect\citeauthoryear{{Cole}, {Percival}, {Peacock}, {Norberg},
  {Baugh}  et~al.}{{Cole} et~al.}{2005}]{Cole_etal2005}
{Cole} S.,  {Percival} W.~J.,  {Peacock} J.~A.,  {Norberg} P.,  {Baugh} C.~M.,
   et~al., 2005, \mn@doi [MNRAS] {10.1111/j.1365-2966.2005.09318.x}, \href
  {http://adsabs.harvard.edu/abs/2005MNRAS.362..505C} {362, 505}

\bibitem[\protect\citeauthoryear{{Collister} \& {Lahav}}{{Collister} \&
  {Lahav}}{2004}]{CollisterLahav_2004}
{Collister} A.~A.,  {Lahav} O.,  2004, \mn@doi [\pasp] {10.1086/383254}, \href
  {https://ui.adsabs.harvard.edu/abs/2004PASP..116..345C} {116, 345}

\bibitem[\protect\citeauthoryear{{Crocce}, {Cabr{\'e}}  \&
  {Gazta{\~n}aga}}{{Crocce} et~al.}{2011}]{CrocceCabreGazta_2011}
{Crocce} M.,  {Cabr{\'e}} A.,   {Gazta{\~n}aga} E.,  2011, \mn@doi [MNRAS]
  {10.1111/j.1365-2966.2011.18393.x}, \href
  {http://adsabs.harvard.edu/abs/2011MNRAS.414..329C} {414, 329}

\bibitem[\protect\citeauthoryear{{Crocce}, {Castander}, {Gazta{\~n}aga},
  {Fosalba}  \& {Carretero}}{{Crocce} et~al.}{2015}]{Crocce_etal2015}
{Crocce} M.,  {Castander} F.~J.,  {Gazta{\~n}aga} E.,  {Fosalba} P.,
  {Carretero} J.,  2015, \mn@doi [MNRAS] {10.1093/mnras/stv1708}, \href
  {https://ui.adsabs.harvard.edu/abs/2015MNRAS.453.1513C} {453, 1513}

\bibitem[\protect\citeauthoryear{{Crocce} et~al.,}{{Crocce}
  et~al.}{2019}]{Crocce_etal2019}
{Crocce} M.,  et~al., 2019, \mn@doi [MNRAS] {10.1093/mnras/sty2522}, \href
  {https://ui.adsabs.harvard.edu/abs/2019MNRAS.482.2807C} {482, 2807}

\bibitem[\protect\citeauthoryear{De~Vicente, S\'anchez  \&
  Sevilla-Noarbe}{De~Vicente et~al.}{2016}]{DeVicente:2015kyp}
De~Vicente J.,  S\'anchez E.,   Sevilla-Noarbe I.,  2016, \mn@doi [Mon. Not.
  Roy. Astron. Soc.] {10.1093/mnras/stw857}, 459, 3078

\bibitem[\protect\citeauthoryear{{Dey} et~al.,}{{Dey}
  et~al.}{2019}]{Dey_etal2019}
{Dey} A.,  et~al., 2019, \mn@doi [\aj] {10.3847/1538-3881/ab089d}, \href
  {https://ui.adsabs.harvard.edu/abs/2019AJ....157..168D} {157, 168}

\bibitem[\protect\citeauthoryear{{Dodelson}}{{Dodelson}}{2003}]{Dodelson_2003}
{Dodelson} S.,  2003, Modern Cosmology.
Academic Press, \mn@doi{10.1016/C2017-0-01943-2}

\bibitem[\protect\citeauthoryear{{Eisenstein}, {Zehavi}, {Hogg}, {Scoccimarro},
  {Blanton}  et~al.}{{Eisenstein} et~al.}{2005}]{Eisenstein_etal2005}
{Eisenstein} D.~J.,  {Zehavi} I.,  {Hogg} D.~W.,  {Scoccimarro} R.,  {Blanton}
  M.~R.,   et~al., 2005, \mn@doi [ApJ] {10.1086/466512}, \href
  {http://adsabs.harvard.edu/abs/2005ApJ...633..560E} {633, 560}

\bibitem[\protect\citeauthoryear{{Erben} et~al.,}{{Erben}
  et~al.}{2009}]{Erben_etal2009}
{Erben} T.,  et~al., 2009, \mn@doi [\aap] {10.1051/0004-6361:200810426}, \href
  {https://ui.adsabs.harvard.edu/abs/2009A&A...493.1197E} {493, 1197}

\bibitem[\protect\citeauthoryear{{Estrada}, {Sefusatti}  \&
  {Frieman}}{{Estrada} et~al.}{2009}]{EstradaSefusattiFrieman2009}
{Estrada} J.,  {Sefusatti} E.,   {Frieman} J.~A.,  2009, \mn@doi [ApJ]
  {10.1088/0004-637X/692/1/265}, \href
  {http://adsabs.harvard.edu/abs/2009ApJ...692..265E} {692, 265}

\bibitem[\protect\citeauthoryear{{Fang}, {Eifler}  \& {Krause}}{{Fang}
  et~al.}{2020}]{cosmolike2020}
{Fang} X.,  {Eifler} T.,   {Krause} E.,  2020, \mn@doi [\mnras]
  {10.1093/mnras/staa1726}, \href
  {https://ui.adsabs.harvard.edu/abs/2020MNRAS.497.2699F} {497, 2699}

\bibitem[\protect\citeauthoryear{{Fosalba}, {Crocce}, {Gazta{\~n}aga}  \&
  {Castander}}{{Fosalba} et~al.}{2015}]{Fosalba_etal2015}
{Fosalba} P.,  {Crocce} M.,  {Gazta{\~n}aga} E.,   {Castander} F.~J.,  2015,
  \mn@doi [MNRAS] {10.1093/mnras/stv138}, \href
  {http://adsabs.harvard.edu/abs/2015MNRAS.448.2987F} {448, 2987}

\bibitem[\protect\citeauthoryear{{Gaia Collaboration} et~al.,}{{Gaia
  Collaboration} et~al.}{2018}]{Gaia_2018}
{Gaia Collaboration} et~al., 2018, \mn@doi [\aap]
  {10.1051/0004-6361/201833051}, \href
  {https://ui.adsabs.harvard.edu/abs/2018A&A...616A...1G} {616, A1}

\bibitem[\protect\citeauthoryear{{Gatti} et~al.,}{{Gatti}
  et~al.}{2018}]{Gatti_etal2018}
{Gatti} M.,  et~al., 2018, \mn@doi [\mnras] {10.1093/mnras/sty466}, \href
  {https://ui.adsabs.harvard.edu/abs/2018MNRAS.477.1664G} {477, 1664}

\bibitem[\protect\citeauthoryear{{Gatti} et~al.,}{{Gatti}
  et~al.}{2022}]{Gatti_etal2022}
{Gatti} M.,  et~al., 2022, \mn@doi [\mnras] {10.1093/mnras/stab3311}, \href
  {https://ui.adsabs.harvard.edu/abs/2022MNRAS.510.1223G} {510, 1223}

\bibitem[\protect\citeauthoryear{Gaztanaga, Cabre  \& Hui}{Gaztanaga
  et~al.}{2009}]{Gaztanaga:2008xz}
Gaztanaga E.,  Cabre A.,   Hui L.,  2009, \mn@doi [Mon. Not. Roy. Astron. Soc.]
  {10.1111/j.1365-2966.2009.15405.x}, 399, 1663

\bibitem[\protect\citeauthoryear{{G{\'o}rski}, {Hivon}, {Banday}, {Wandelt},
  {Hansen}, {Reinecke}  \& {Bartelmann}}{{G{\'o}rski} et~al.}{2005}]{Healpix}
{G{\'o}rski} K.~M.,  {Hivon} E.,  {Banday} A.~J.,  {Wandelt} B.~D.,  {Hansen}
  F.~K.,  {Reinecke} M.,   {Bartelmann} M.,  2005, \mn@doi [\apj]
  {10.1086/427976}, \href
  {https://ui.adsabs.harvard.edu/abs/2005ApJ...622..759G} {622, 759}

\bibitem[\protect\citeauthoryear{{Ho} et~al.,}{{Ho} et~al.}{2012}]{Ho_etal2012}
{Ho} S.,  et~al., 2012, \mn@doi [\apj] {10.1088/0004-637X/761/1/14}, \href
  {https://ui.adsabs.harvard.edu/abs/2012ApJ...761...14H} {761, 14}

\bibitem[\protect\citeauthoryear{{Hu} \& {Sugiyama}}{{Hu} \&
  {Sugiyama}}{1996}]{HuSugiyama1996}
{Hu} W.,  {Sugiyama} N.,  1996, \mn@doi [ApJ] {10.1086/177989}, \href
  {http://adsabs.harvard.edu/abs/1996ApJ...471..542H} {471, 542}

\bibitem[\protect\citeauthoryear{{Hu}, {Sugiyama}  \& {Silk}}{{Hu}
  et~al.}{1997}]{HuSugiyamaSilk1997}
{Hu} W.,  {Sugiyama} N.,   {Silk} J.,  1997, \mn@doi [Nature]
  {10.1038/386037a0}, \href {http://adsabs.harvard.edu/abs/1997Natur.386...37H}
  {386, 37}

\bibitem[\protect\citeauthoryear{{H{\"u}tsi}}{{H{\"u}tsi}}{2010}]{Hutsi2010}
{H{\"u}tsi} G.,  2010, \mn@doi [\mnras] {10.1111/j.1365-2966.2009.15824.x},
  \href {http://adsabs.harvard.edu/abs/2010MNRAS.401.2477H} {401, 2477}

\bibitem[\protect\citeauthoryear{{Ilbert} et~al.,}{{Ilbert}
  et~al.}{2006}]{Ilbert_etal2006}
{Ilbert} O.,  et~al., 2006, \mn@doi [\aap] {10.1051/0004-6361:20065138}, \href
  {https://ui.adsabs.harvard.edu/abs/2006A&A...457..841I} {457, 841}

\bibitem[\protect\citeauthoryear{{Ishikawa}, {Sunayama}, {Nishizawa},
  {Miyatake}  \& {Nishimichi}}{{Ishikawa} et~al.}{2023}]{Ishikawa_etal2023}
{Ishikawa} K.,  {Sunayama} T.,  {Nishizawa} A.~J.,  {Miyatake} H.,
  {Nishimichi} T.,  2023, \mn@doi [arXiv e-prints] {10.48550/arXiv.2306.01696},
  \href {https://ui.adsabs.harvard.edu/abs/2023arXiv230601696I} {p.
  arXiv:2306.01696}

\bibitem[\protect\citeauthoryear{Ivanov \& Sibiryakov}{Ivanov \&
  Sibiryakov}{2018}]{Ivanov:2018gjr}
Ivanov M.~M.,  Sibiryakov S.,  2018, \mn@doi [JCAP]
  {10.1088/1475-7516/2018/07/053}, 07, 053

\bibitem[\protect\citeauthoryear{{Kaiser}}{{Kaiser}}{1987}]{Kaiser87}
{Kaiser} N.,  1987, \mn@doi [MNRAS] {10.1093/mnras/227.1.1}, \href
  {https://ui.adsabs.harvard.edu/abs/1987MNRAS.227....1K} {227, 1}

\bibitem[\protect\citeauthoryear{{Kazin}, {Koda}, {Blake}, {Padmanabhan},
  {Brough}  et~al.}{{Kazin} et~al.}{2014}]{Kazin_etal2014}
{Kazin} E.~A.,  {Koda} J.,  {Blake} C.,  {Padmanabhan} N.,  {Brough} S.,
  et~al., 2014, \mn@doi [MNRAS] {10.1093/mnras/stu778}, \href
  {http://adsabs.harvard.edu/abs/2014MNRAS.441.3524K} {441, 3524}

\bibitem[\protect\citeauthoryear{{Krause} \& {Eifler}}{{Krause} \&
  {Eifler}}{2017}]{cosmolike}
{Krause} E.,  {Eifler} T.,  2017, \mn@doi [\mnras] {10.1093/mnras/stx1261},
  \href {https://ui.adsabs.harvard.edu/abs/2017MNRAS.470.2100K} {470, 2100}

\bibitem[\protect\citeauthoryear{{Landy} \& {Szalay}}{{Landy} \&
  {Szalay}}{1993}]{LandySzalay_1993}
{Landy} S.~D.,  {Szalay} A.~S.,  1993, \mn@doi [\apj] {10.1086/172900}, \href
  {http://adsabs.harvard.edu/abs/1993ApJ...412...64L} {412, 64}

\bibitem[\protect\citeauthoryear{{Lang}, {Hogg}  \& {Mykytyn}}{{Lang}
  et~al.}{2016}]{Lang_etal2016}
{Lang} D.,  {Hogg} D.~W.,   {Mykytyn} D.,  2016, {The Tractor: Probabilistic
  astronomical source detection and measurement}, Astrophysics Source Code
  Library, record ascl:1604.008 (\mn@eprint {ascl} {1604.008})

\bibitem[\protect\citeauthoryear{{Lee} \& {Seung}}{{Lee} \&
  {Seung}}{1999}]{LeeSeung1999}
{Lee} D.~D.,  {Seung} H.~S.,  1999, \mn@doi [\nat] {10.1038/44565}, \href
  {https://ui.adsabs.harvard.edu/abs/1999Natur.401..788L} {401, 788}

\bibitem[\protect\citeauthoryear{Lewis \& Challinor}{Lewis \&
  Challinor}{2007}]{Lewis:2007kz}
Lewis A.,  Challinor A.,  2007, \mn@doi [Phys. Rev. D]
  {10.1103/PhysRevD.76.083005}, 76, 083005

\bibitem[\protect\citeauthoryear{{Li}, {Napolitano}, {Roy}, {Tortora}, {La
  Barbera}, {Sonnenfeld}, {Qiu}  \& {Liu}}{{Li}
  et~al.}{2022}]{LiNapolitano_etal2022}
{Li} R.,  {Napolitano} N.~R.,  {Roy} N.,  {Tortora} C.,  {La Barbera} F.,
  {Sonnenfeld} A.,  {Qiu} C.,   {Liu} S.,  2022, \mn@doi [\apj]
  {10.3847/1538-4357/ac5ea0}, \href
  {https://ui.adsabs.harvard.edu/abs/2022ApJ...929..152L} {929, 152}

\bibitem[\protect\citeauthoryear{{Lima}, {Cunha}, {Oyaizu}, {Frieman}, {Lin}
  \& {Sheldon}}{{Lima} et~al.}{2008}]{Lima_etal2008}
{Lima} M.,  {Cunha} C.~E.,  {Oyaizu} H.,  {Frieman} J.,  {Lin} H.,   {Sheldon}
  E.~S.,  2008, \mn@doi [\mnras] {10.1111/j.1365-2966.2008.13510.x}, \href
  {https://ui.adsabs.harvard.edu/abs/2008MNRAS.390..118L} {390, 118}

\bibitem[\protect\citeauthoryear{{Matthews} \& {Newman}}{{Matthews} \&
  {Newman}}{2010}]{MatthewsNewman_2010}
{Matthews} D.~J.,  {Newman} J.~A.,  2010, \mn@doi [\apj]
  {10.1088/0004-637X/721/1/456}, \href
  {https://ui.adsabs.harvard.edu/abs/2010ApJ...721..456M} {721, 456}

\bibitem[\protect\citeauthoryear{{McQuinn} \& {White}}{{McQuinn} \&
  {White}}{2013}]{McQuinnWhite_2013}
{McQuinn} M.,  {White} M.,  2013, \mn@doi [\mnras] {10.1093/mnras/stt914},
  \href {https://ui.adsabs.harvard.edu/abs/2013MNRAS.433.2857M} {433, 2857}

\bibitem[\protect\citeauthoryear{{Mena-Fern{\'a}ndez}
  et~al.,}{{Mena-Fern{\'a}ndez} et~al.}{2024}]{DESY6_BAOsample}
{Mena-Fern{\'a}ndez} J.,  et~al., 2024, \mn@doi [arXiv e-prints]
  {10.48550/arXiv.2402.10697}, \href
  {https://ui.adsabs.harvard.edu/abs/2024arXiv240210697M} {p. arXiv:2402.10697}

\bibitem[\protect\citeauthoryear{{M{\'e}nard}, {Scranton}, {Schmidt},
  {Morrison}, {Jeong}, {Budavari}  \& {Rahman}}{{M{\'e}nard}
  et~al.}{2013}]{Menard_etal2013}
{M{\'e}nard} B.,  {Scranton} R.,  {Schmidt} S.,  {Morrison} C.,  {Jeong} D.,
  {Budavari} T.,   {Rahman} M.,  2013, \mn@doi [arXiv e-prints]
  {10.48550/arXiv.1303.4722}, \href
  {https://ui.adsabs.harvard.edu/abs/2013arXiv1303.4722M} {p. arXiv:1303.4722}

\bibitem[\protect\citeauthoryear{{Moessner} \& {Jain}}{{Moessner} \&
  {Jain}}{1998}]{MoessnerJain1998}
{Moessner} R.,  {Jain} B.,  1998, \mn@doi [\mnras]
  {10.1046/j.1365-8711.1998.01378.x10.1111/j.1365-8711.1998.01378.x}, \href
  {https://ui.adsabs.harvard.edu/abs/1998MNRAS.294L..18M} {294, L18}

\bibitem[\protect\citeauthoryear{{Mohammad} \& {Percival}}{{Mohammad} \&
  {Percival}}{2022}]{Mohammad_Percival2022}
{Mohammad} F.~G.,  {Percival} W.~J.,  2022, \mn@doi [\mnras]
  {10.1093/mnras/stac1458}, \href
  {https://ui.adsabs.harvard.edu/abs/2022MNRAS.514.1289M} {514, 1289}

\bibitem[\protect\citeauthoryear{{Moon}, {Valcin}, {Rashkovetskyi}, {Saulder},
  {Aguilar}  et~al.}{{Moon} et~al.}{2023}]{DESI_BAO_2023}
{Moon} J.,  {Valcin} D.,  {Rashkovetskyi} M.,  {Saulder} C.,  {Aguilar} J.~N.,
   et~al., 2023, \mn@doi [arXiv e-prints] {10.48550/arXiv.2304.08427}, \href
  {https://ui.adsabs.harvard.edu/abs/2023arXiv230408427M} {p. arXiv:2304.08427}

\bibitem[\protect\citeauthoryear{{Morrison}, {Hildebrandt}, {Schmidt},
  {Baldry}, {Bilicki}, {Choi}, {Erben}  \& {Schneider}}{{Morrison}
  et~al.}{2017}]{Morrison_etal2017}
{Morrison} C.~B.,  {Hildebrandt} H.,  {Schmidt} S.~J.,  {Baldry} I.~K.,
  {Bilicki} M.,  {Choi} A.,  {Erben} T.,   {Schneider} P.,  2017, \mn@doi
  [\mnras] {10.1093/mnras/stx342}, \href
  {https://ui.adsabs.harvard.edu/abs/2017MNRAS.467.3576M} {467, 3576}

\bibitem[\protect\citeauthoryear{{Newman}}{{Newman}}{2008}]{Newman_2008}
{Newman} J.~A.,  2008, \mn@doi [\apj] {10.1086/589982}, \href
  {https://ui.adsabs.harvard.edu/abs/2008ApJ...684...88N} {684, 88}

\bibitem[\protect\citeauthoryear{{Newman} \& {Gruen}}{{Newman} \&
  {Gruen}}{2022}]{NewmanGruen_2022}
{Newman} J.~A.,  {Gruen} D.,  2022, \mn@doi [\araa]
  {10.1146/annurev-astro-032122-014611}, \href
  {https://ui.adsabs.harvard.edu/abs/2022ARA&A..60..363N} {60, 363}

\bibitem[\protect\citeauthoryear{{Norberg}, {Baugh}, {Gazta{\~n}aga}  \&
  {Croton}}{{Norberg} et~al.}{2009}]{Norberg09}
{Norberg} P.,  {Baugh} C.~M.,  {Gazta{\~n}aga} E.,   {Croton} D.~J.,  2009,
  \mn@doi [\mnras] {10.1111/j.1365-2966.2009.14389.x}, \href
  {https://ui.adsabs.harvard.edu/abs/2009MNRAS.396...19N} {396, 19}

\bibitem[\protect\citeauthoryear{{Padmanabhan}, {Schlegel}, {Seljak},
  {Makarov}, {Bahcall}  et~al.}{{Padmanabhan}
  et~al.}{2007}]{Padmanabhan_etal2007}
{Padmanabhan} N.,  {Schlegel} D.~J.,  {Seljak} U.,  {Makarov} A.,  {Bahcall}
  N.~A.,   et~al., 2007, \mn@doi [MNRAS] {10.1111/j.1365-2966.2007.11593.x},
  \href {http://adsabs.harvard.edu/abs/2007MNRAS.378..852P} {378, 852}

\bibitem[\protect\citeauthoryear{{Peebles} \& {Yu}}{{Peebles} \&
  {Yu}}{1970}]{PeeblesYu1970}
{Peebles} P.~J.~E.,  {Yu} J.~T.,  1970, \mn@doi [ApJ] {10.1086/150713}, \href
  {http://adsabs.harvard.edu/abs/1970ApJ...162..815P} {162, 815}

\bibitem[\protect\citeauthoryear{{Peng}, {Xu}, {Zhang}, {Chen}  \& {Yu}}{{Peng}
  et~al.}{2022}]{Peng_etal2022}
{Peng} H.,  {Xu} H.,  {Zhang} L.,  {Chen} Z.,   {Yu} Y.,  2022, \mn@doi
  [\mnras] {10.1093/mnras/stac2713}, \href
  {https://ui.adsabs.harvard.edu/abs/2022MNRAS.516.6210P} {516, 6210}

\bibitem[\protect\citeauthoryear{{Percival}, {Reid}, {Eisenstein}, {Bahcall},
  {Budavari}  et~al.}{{Percival} et~al.}{2010}]{Percival_etal2010}
{Percival} W.~J.,  {Reid} B.~A.,  {Eisenstein} D.~J.,  {Bahcall} N.~A.,
  {Budavari} T.,   et~al., 2010, \mn@doi [MNRAS]
  {10.1111/j.1365-2966.2009.15812.x}, \href
  {http://adsabs.harvard.edu/abs/2010MNRAS.401.2148P} {401, 2148}

\bibitem[\protect\citeauthoryear{{Planck Collaboration}, {Aghanim}, {Akrami},
  {Ashdown}, {Aumont}  et~al.}{{Planck Collaboration}
  et~al.}{2020}]{Planck2020}
{Planck Collaboration} {Aghanim} N.,  {Akrami} Y.,  {Ashdown} M.,  {Aumont} J.,
    et~al., 2020, \mn@doi [\aap] {10.1051/0004-6361/201833910}, \href
  {https://ui.adsabs.harvard.edu/abs/2020A&A...641A...6P} {641, A6}

\bibitem[\protect\citeauthoryear{{Rau} et~al.,}{{Rau}
  et~al.}{2023}]{Rau_etal2023}
{Rau} M.~M.,  et~al., 2023, \mn@doi [\mnras] {10.1093/mnras/stad1962}, \href
  {https://ui.adsabs.harvard.edu/abs/2023MNRAS.524.5109R} {524, 5109}

\bibitem[\protect\citeauthoryear{{Rezaie}, {Seo}, {Ross}  \&
  {Bunescu}}{{Rezaie} et~al.}{2020}]{Rezaie_etal2020}
{Rezaie} M.,  {Seo} H.-J.,  {Ross} A.~J.,   {Bunescu} R.~C.,  2020, \mn@doi
  [\mnras] {10.1093/mnras/staa1231}, \href
  {https://ui.adsabs.harvard.edu/abs/2020MNRAS.495.1613R} {495, 1613}

\bibitem[\protect\citeauthoryear{{Rodr{\'\i}guez-Monroy}
  et~al.,}{{Rodr{\'\i}guez-Monroy} et~al.}{2022}]{Rodriguez-Monroy_etal2022}
{Rodr{\'\i}guez-Monroy} M.,  et~al., 2022, \mn@doi [\mnras]
  {10.1093/mnras/stac104}, \href
  {https://ui.adsabs.harvard.edu/abs/2022MNRAS.511.2665R} {511, 2665}

\bibitem[\protect\citeauthoryear{{Ross} et~al.,}{{Ross}
  et~al.}{2011}]{Ross_etal2011}
{Ross} A.~J.,  et~al., 2011, \mn@doi [\mnras]
  {10.1111/j.1365-2966.2011.19351.x}, \href
  {https://ui.adsabs.harvard.edu/abs/2011MNRAS.417.1350R} {417, 1350}

\bibitem[\protect\citeauthoryear{{Ross}, {Samushia}, {Howlett}, {Percival},
  {Burden}  \& {Manera}}{{Ross} et~al.}{2015}]{Ross_etal2015}
{Ross} A.~J.,  {Samushia} L.,  {Howlett} C.,  {Percival} W.~J.,  {Burden} A.,
  {Manera} M.,  2015, \mn@doi [MNRAS] {10.1093/mnras/stv154}, \href
  {http://adsabs.harvard.edu/abs/2015MNRAS.449..835R} {449, 835}

\bibitem[\protect\citeauthoryear{{Ross} et~al.,}{{Ross}
  et~al.}{2017a}]{Ross_etal2017}
{Ross} A.~J.,  et~al., 2017a, \mn@doi [\mnras] {10.1093/mnras/stw2372}, \href
  {https://ui.adsabs.harvard.edu/abs/2017MNRAS.464.1168R} {464, 1168}

\bibitem[\protect\citeauthoryear{Ross et~al.}{Ross
  et~al.}{2017b}]{Ross:2017emc}
Ross A.~J.,  et~al., 2017b, \mn@doi [Mon. Not. Roy. Astron. Soc.]
  {10.1093/mnras/stx2120}, 472, 4456

\bibitem[\protect\citeauthoryear{Sadeh, Abdalla  \& Lahav}{Sadeh
  et~al.}{2016}]{Sadeh:2015lsa}
Sadeh I.,  Abdalla F.~B.,   Lahav O.,  2016, \mn@doi [Publ. Astron. Soc. Pac.]
  {10.1088/1538-3873/128/968/104502}, 128, 104502

\bibitem[\protect\citeauthoryear{{Salazar-Albornoz} et~al.,}{{Salazar-Albornoz}
  et~al.}{2017}]{Salazar-Albornoz:2016psd}
{Salazar-Albornoz} S.,  et~al., 2017, \mn@doi [MNRAS] {10.1093/mnras/stx633},
  \href {http://adsabs.harvard.edu/abs/2017MNRAS.468.2938S} {468, 2938}

\bibitem[\protect\citeauthoryear{{Salvato}, {Ilbert}  \& {Hoyle}}{{Salvato}
  et~al.}{2019}]{Salvato_etal2019}
{Salvato} M.,  {Ilbert} O.,   {Hoyle} B.,  2019, \mn@doi [Nature Astronomy]
  {10.1038/s41550-018-0478-0}, \href
  {https://ui.adsabs.harvard.edu/abs/2019NatAs...3..212S} {3, 212}

\bibitem[\protect\citeauthoryear{{Schlafly} \& {Finkbeiner}}{{Schlafly} \&
  {Finkbeiner}}{2011}]{Schlafly_etal2011}
{Schlafly} E.~F.,  {Finkbeiner} D.~P.,  2011, \mn@doi [\apj]
  {10.1088/0004-637X/737/2/103}, \href
  {https://ui.adsabs.harvard.edu/abs/2011ApJ...737..103S} {737, 103}

\bibitem[\protect\citeauthoryear{{Schlegel}, {Finkbeiner}  \&
  {Davis}}{{Schlegel} et~al.}{1998}]{Schlegel_etal1998}
{Schlegel} D.~J.,  {Finkbeiner} D.~P.,   {Davis} M.,  1998, \mn@doi [\apj]
  {10.1086/305772}, \href
  {https://ui.adsabs.harvard.edu/abs/1998ApJ...500..525S} {500, 525}

\bibitem[\protect\citeauthoryear{{Schneider}, {Knox}, {Zhan}  \&
  {Connolly}}{{Schneider} et~al.}{2006}]{Schneider_etal2006}
{Schneider} M.,  {Knox} L.,  {Zhan} H.,   {Connolly} A.,  2006, \mn@doi [\apj]
  {10.1086/507675}, \href
  {https://ui.adsabs.harvard.edu/abs/2006ApJ...651...14S} {651, 14}

\bibitem[\protect\citeauthoryear{{Scranton} et~al.,}{{Scranton}
  et~al.}{2002}]{Scranton_etal2002}
{Scranton} R.,  et~al., 2002, \mn@doi [\apj] {10.1086/342786}, \href
  {https://ui.adsabs.harvard.edu/abs/2002ApJ...579...48S} {579, 48}

\bibitem[\protect\citeauthoryear{{Seo} \& {Eisenstein}}{{Seo} \&
  {Eisenstein}}{2003}]{SeoEisenstein_2003}
{Seo} H.-J.,  {Eisenstein} D.~J.,  2003, \mn@doi [ApJ] {10.1086/379122}, \href
  {https://ui.adsabs.harvard.edu/abs/2003ApJ...598..720S} {598, 720}

\bibitem[\protect\citeauthoryear{{Seo}, {Ho}, {White}, {Cuesta}, {Ross}
  et~al.}{{Seo} et~al.}{2012}]{Seo_etal2012}
{Seo} H.-J.,  {Ho} S.,  {White} M.,  {Cuesta} A.~J.,  {Ross} A.~J.,   et~al.,
  2012, \mn@doi [\apj] {10.1088/0004-637X/761/1/13}, \href
  {http://adsabs.harvard.edu/abs/2012ApJ...761...13S} {761, 13}

\bibitem[\protect\citeauthoryear{{Simon}}{{Simon}}{2007}]{Simon2007}
{Simon} P.,  2007, \mn@doi [\aap] {10.1051/0004-6361:20066352}, \href
  {https://ui.adsabs.harvard.edu/abs/2007A&A...473..711S} {473, 711}

\bibitem[\protect\citeauthoryear{{Sinha} \& {Garrison}}{{Sinha} \&
  {Garrison}}{2020}]{SinhaGarrison_2020}
{Sinha} M.,  {Garrison} L.~H.,  2020, \mn@doi [\mnras] {10.1093/mnras/stz3157},
  \href {https://ui.adsabs.harvard.edu/abs/2020MNRAS.491.3022S} {491, 3022}

\bibitem[\protect\citeauthoryear{{Sunyaev} \& {Zeldovich}}{{Sunyaev} \&
  {Zeldovich}}{1970}]{SunyaevZeldovich1970}
{Sunyaev} R.~A.,  {Zeldovich} Y.~B.,  1970, \mn@doi [Astrophysics and Space
  Science] {10.1007/BF00653471}, \href
  {http://adsabs.harvard.edu/abs/1970Ap%26SS...7....3S} {7, 3}

\bibitem[\protect\citeauthoryear{{Troxel}, {Krause}, {Chang}, {Eifler},
  {Friedrich}, {Gruen}, {MacCrann}  et~al.}{{Troxel}
  et~al.}{2018}]{cosmolike_mask}
{Troxel} M.~A.,  {Krause} E.,  {Chang} C.,  {Eifler} T.~F.,  {Friedrich} O.,
  {Gruen} D.,  {MacCrann} N.,   et~al., 2018, \mn@doi [\mnras]
  {10.1093/mnras/sty1889}, \href
  {https://ui.adsabs.harvard.edu/abs/2018MNRAS.479.4998T} {479, 4998}

\bibitem[\protect\citeauthoryear{{Weinberg}, {Mortonson}, {Eisenstein},
  {Hirata}, {Riess}  \& {Rozo}}{{Weinberg}
  et~al.}{2013}]{WeinbergMortonson_etal2013}
{Weinberg} D.~H.,  {Mortonson} M.~J.,  {Eisenstein} D.~J.,  {Hirata} C.,
  {Riess} A.~G.,   {Rozo} E.,  2013, \mn@doi [\physrep]
  {10.1016/j.physrep.2013.05.001}, \href
  {http://adsabs.harvard.edu/abs/2013PhR...530...87W} {530, 87}

\bibitem[\protect\citeauthoryear{{Xu} et~al.,}{{Xu} et~al.}{2023}]{XU2023}
{Xu} H.,  et~al., 2023, \mn@doi [MNRAS] {10.1093/mnras/stad136}, \href
  {https://ui.adsabs.harvard.edu/abs/2023MNRAS.520..161X} {520, 161}

\bibitem[\protect\citeauthoryear{{Yang} et~al.,}{{Yang}
  et~al.}{2021}]{Yang_etal2021}
{Yang} X.,  et~al., 2021, \mn@doi [\apj] {10.3847/1538-4357/abddb2}, \href
  {https://ui.adsabs.harvard.edu/abs/2021ApJ...909..143Y} {909, 143}

\bibitem[\protect\citeauthoryear{Zhan, Knox  \& Tyson}{Zhan
  et~al.}{2009}]{Zhan:2008jh}
Zhan H.,  Knox L.,   Tyson J.~A.,  2009, \mn@doi [\apj]
  {10.1088/0004-637X/690/1/923}, 690, 923

\bibitem[\protect\citeauthoryear{{Zhang}, {Pen}  \& {Bernstein}}{{Zhang}
  et~al.}{2010}]{Zhang_etal2010}
{Zhang} P.,  {Pen} U.-L.,   {Bernstein} G.,  2010, \mn@doi [\mnras]
  {10.1111/j.1365-2966.2010.16445.x}, \href
  {https://ui.adsabs.harvard.edu/abs/2010MNRAS.405..359Z} {405, 359}

\bibitem[\protect\citeauthoryear{{Zhang}, {Yu}  \& {Zhang}}{{Zhang}
  et~al.}{2017}]{Zhang_etal2017}
{Zhang} L.,  {Yu} Y.,   {Zhang} P.,  2017, \mn@doi [\apj]
  {10.3847/1538-4357/aa8c72}, \href
  {https://ui.adsabs.harvard.edu/abs/2017ApJ...848...44Z} {848, 44}

\bibitem[\protect\citeauthoryear{{Zhou} et~al.,}{{Zhou}
  et~al.}{2021}]{Zhou_etal2021}
{Zhou} R.,  et~al., 2021, \mn@doi [\mnras] {10.1093/mnras/staa3764}, \href
  {https://ui.adsabs.harvard.edu/abs/2021MNRAS.501.3309Z} {501, 3309}

\bibitem[\protect\citeauthoryear{{de Simoni}, {Sobreira}, {Carnero}, {Ross},
  {Camacho}  et~al.}{{de Simoni} et~al.}{2013}]{deSimoni_etal2013}
{de Simoni} F.,  {Sobreira} F.,  {Carnero} A.,  {Ross} A.~J.,  {Camacho} H.~O.,
    et~al., 2013, \mn@doi [\mnras] {10.1093/mnras/stt1496}, \href
  {http://adsabs.harvard.edu/abs/2013MNRAS.435.3017D} {435, 3017}

\bibitem[\protect\citeauthoryear{{von Wietersheim-Kramsta} et~al.,}{{von
  Wietersheim-Kramsta} et~al.}{2021}]{vonWietersheim-Kramsta2021}
{von Wietersheim-Kramsta} M.,  et~al., 2021, \mn@doi [\mnras]
  {10.1093/mnras/stab1000}, \href
  {https://ui.adsabs.harvard.edu/abs/2021MNRAS.504.1452V} {504, 1452}

\makeatother
\end{thebibliography}









\bsp	
\label{lastpage}
\end{document}